\documentclass[final,3p,times,twocolumn,nonatbib]{elsarticle}
\usepackage[utf8]{inputenc}
\usepackage[american]{babel}
\usepackage{csquotes}
\usepackage[backend=biber,style=apa]{biblatex}
\usepackage[T1]{fontenc}
\usepackage[utf8]{inputenc}
\usepackage{textcomp}
\usepackage{newtxtext,newtxmath}
\usepackage{graphicx,booktabs,array,tabularx,longtable}
\usepackage{microtype}
\usepackage{xurl}
\usepackage[dvipsnames]{xcolor}
\usepackage[colorlinks=true,linkcolor=MidnightBlue,citecolor=MidnightBlue,urlcolor=MidnightBlue,breaklinks=true]{hyperref}
\usepackage{caption}
\newcolumntype{L}[1]{>{\raggedright\arraybackslash}p{#1}}
\newcolumntype{Y}{>{\raggedright\arraybackslash}X}
\usepackage{flushend}
\journal{Preprint}

\makeatletter
\def\ps@pprintTitle{\let\@oddhead\@empty\let\@evenhead\@empty\def\@oddfoot{\footnotesize\itshape Preprint\hfill 15 September 2026}\let\@evenfoot\@oddfoot}
\makeatother
\hypersetup{pdftitle={Ageing, Digital Literacy, and Interaction Modality in Immersive Virtual Reality: Psychomotor Performance, Cognitive Flexibility, and Their Processing-Speed Association},pdfauthor={Panagiotis Kourtesis; Katerina Denaxa; Lydia Asimakopoulou; Petros Roussos; Maria Roussou}}
\begin{document}
\begin{frontmatter}
\title{Ageing, Digital Literacy, and Interaction Modality in Immersive Virtual Reality: Psychomotor Performance, Cognitive Flexibility, and Their Processing-Speed Association}
\author[aff1,aff2,aff3,aff4]{\href{https://orcid.org/0000-0002-2914-1064}{Panagiotis Kourtesis}\corref{cor1}}
\author[aff2]{\href{https://orcid.org/0000-0002-8704-2697}{Katerina Denaxa}}
\author[aff2]{\href{https://orcid.org/0009-0004-6770-8852}{Lydia Asimakopoulou}}
\author[aff2]{\href{https://orcid.org/0000-0003-1465-2117}{Petros Roussos}}
\author[aff1]{\href{https://orcid.org/0000-0002-2826-162X}{Maria Roussou}}
\address[aff1]{Department of Informatics and Telecommunications, National and Kapodistrian University of Athens, Athens, Greece; mroussou@di.uoa.gr (M.R.)}
\address[aff2]{Department of Psychology, National and Kapodistrian University of Athens, Athens, Greece; pkourtesis@psych.uoa.gr (P.K.); adenaxa@psych.uoa.gr (K.D.); lydiaas@psych.uoa.gr (L.A.); roussosp@psych.uoa.gr (P.R.)}
\address[aff3]{Department of Psychology, The American College of Greece, Athens, Greece}
\address[aff4]{Department of Psychology, University of Edinburgh, Edinburgh, United Kingdom}
\cortext[cor1]{Correspondence: \href{mailto:pkourtesis@psych.uoa.gr}{pkourtesis@psych.uoa.gr}}
\begin{abstract}
Extended reality (XR) increasingly supports training and cognitive assessment, yet the age sensitivity of its interaction techniques is unclear. This study examined age, digital literacy, and interaction modality as correlates of psychomotor and cognitive-flexibility performance in immersive virtual reality (VR). Two hundred and two adults (19--90 years) completed a five-mode Fitts' law task (eye-gaze, head-gaze, controller ray-casting, virtual finger, and controller direct touch), the Trail Making Test in VR (TMT-VR), and a digital-literacy questionnaire. Age was associated with slower task times across modes, but controller direct touch carried the steepest relative age gradient yet remained among the fastest in absolute terms; technique altered relative age sensitivity without determining absolute efficiency. Higher digital literacy was associated with faster TMT-VR completion but not Fitts task times. A Fitts-derived speed score predicted TMT-VR performance beyond age and partly accounted for its age association, consistent with shared processing-speed variance; an exploratory full-battery extension confined the Fitts-score association to completion-time and error-adjusted-time indices and the digital-literacy association to error-adjusted-time indices, not wrong-target errors, mean selection distance, or relative Part B indices. Age-inclusive XR assessment should standardise interaction modality, evaluate rather than exclude mid-air direct selection, and interpret cognitive scores alongside digital literacy and psychomotor speed.
\end{abstract}
\begin{keyword}
immersive virtual reality \sep  ageing \sep  digital literacy \sep  interaction modality \sep  Fitts' law \sep  cognitive flexibility \sep  processing speed \sep  human--computer interaction
\end{keyword}
\end{frontmatter}

\section{Introduction}\label{sec:1}

\subsection{Interaction techniques and ageing in immersive VR}\label{sec:1.1}

Immersive virtual reality (VR) has matured into a viable medium for skill training, neuropsychological assessment, and everyday applications, with head-mounted displays (HMDs) now supporting several distinct selection techniques, namely eye-gaze, head-gaze, controller-based ray-casting, and direct touch using a virtual hand or controller \parencite{ref1,ref2}. The proliferation of these techniques raises a practical question for human--computer interaction (HCI), namely which technique should be preferred, for whom, and for which task demands. This question becomes pressing as VR user populations broaden to include middle-aged and older adults, whose psychomotor and cognitive characteristics differ systematically from those of the young adults on whom most extended-reality (XR) interaction research has been conducted \parencite{ref3,ref4}.

Fitts' law \parencite{ref5} remains the canonical framework for quantifying pointing performance, relating movement time (MT) to an index of difficulty (ID) determined by target width and distance \parencite{ref6,ref7}. In VR, Fitts-type tasks have been used to compare interaction techniques in young adults, indicating advantages for ray-casting in speed and for direct manipulation in intuitiveness, as well as costs associated with the absence of haptic feedback when pointing with a bare virtual hand \parencite{ref8}. Comparisons of eye-gaze, head-gaze, and controller pointing for target selection in HMDs, again in young adults, have found eye-gaze pointing faster than head pointing and comparable to controller pointing \parencite{ref9,ref10}. However, the interaction between selection technique and user age has received less direct attention. The non-immersive HCI literature indicates that age effects on pointing depend on the input device: age costs were larger for touchpad than mouse input \parencite{ref11}, whereas direct touch input attenuated the age differences observed with the mouse \parencite{ref12}.

In immersive VR, age-comparative evidence exists but is sparse: a CAVE-type comparison with mouse and touchscreen input found older adults slower and more error-prone overall and concluded that the tested virtual environment suited scene viewing better than fine manipulation \parencite{ref13}, and a recent HMD study found poorer predictive (feedforward) control of the final reaching position in older adults for long stereoscopic reaches \parencite{ref14}. A recent HMD study of selection and manipulation tasks likewise reported age-related difficulties, particularly with targets outside the field of view and in spatial perception, and derived interface recommendations for older users \parencite{ref15}. In the non-immersive touch literature, touchscreen input reduced the young--old performance gap relative to the mouse \parencite{ref16}. Ageing is, moreover, associated with slower and more variable aimed movements and with disproportionate slowing under higher task difficulty \parencite{ref17,ref18,ref19}. To our knowledge, however, no study has compared age gradients across the selection techniques available within a single modern HMD and interaction configuration. Whether age effects are uniform across XR selection techniques, or whether some techniques carry shallower age gradients, therefore remains an open question that bears directly on age-inclusive XR design; the present within-HMD comparison addresses this gap.

\subsection{Cognitive ageing, the Trail Making Test, and processing speed}\label{sec:1.2}

Immersive VR is also increasingly used for cognitive assessment, and a parallel literature concerns cognitive ageing. The Trail Making Test (TMT) \parencite{ref20} is multicomponent in both parts: Part A draws on visual search and visuomotor speed, and Part B additionally loads working memory, alternation, and set maintenance, with derived difference and ratio scores showing limited convergence as pure executive measures \parencite{ref21,ref22,ref23,ref24,ref25}. The TMT-VR, an immersive adaptation validated across interaction modes in young adults and examined in clinical populations \parencite{ref26,ref27,ref28}, permits the psychomotor and cognitive strands to be examined within the same testing session and measurement environment. The two tasks differ in their demands: in the Fitts task attention is guided to a single highlighted target, so performance depends chiefly on psychomotor speed, whereas in the TMT-VR the participant must search for each successive target, so attention, accuracy, and speed all contribute to performance \parencite{ref26,ref27}. A long-standing theoretical position holds that much of cognitive ageing reflects a general slowing of processing speed rather than domain-specific decline \parencite{ref29,ref30,ref31,ref32,ref33,ref34,ref35,ref36}. This account generates two testable predictions in the present design: first, the age association with TMT-VR performance should be largely speed-general, such that Part B is not disproportionately affected once general slowing is accounted for; second, an individual-differences index of psychomotor speed derived from the Fitts task should predict TMT-VR performance over and above chronological age, and basic speed (Part A) should statistically account for a substantial share of the age effect on Part B.

\subsection{Digital literacy}\label{sec:1.3}

Digital literacy constitutes a further individual difference of applied importance. Older adults vary widely in technology experience, adoption, and self-efficacy \parencite{ref37,ref38}. Whether technology-mediated assessment scores partly reflect digital familiarity rather than the target construct remains an open empirical question \parencite{ref37,ref39}. In computerised neuropsychological testing, self-reported computer familiarity has been associated with better scores on speeded tests of visual scanning and attention \parencite{ref40,ref41}, although among older adults the association was no stronger for computerised than for paper-and-pencil measures \parencite{ref42}, suggesting that technology experience may partly index general cognitive or motivational factors rather than test-specific familiarity. Consistent with this reading, a recent meta-analysis found everyday use of digital technology to be associated with a lower risk of cognitive impairment in later life \parencite{ref43}. In immersive VR, technological competence has been identified as a pre-condition for the effective use of head-mounted displays \parencite{ref39}, whereas a direct comparison of immersive and non-immersive versions of working-memory and psychomotor tasks found performance in VR to be largely independent of computing and gaming experience, unlike its non-immersive counterpart \parencite{ref44}. In particular, it is unresolved whether digital literacy is associated with psychomotor performance in VR, moderates the age association, or relates selectively to cognitive task performance.

\subsection{Aims and hypotheses}\label{sec:1.4}

The present study addressed these questions in a single design combining a five-mode Fitts' law task and the TMT-VR. Specifically, it had four aims: (1) to compare interaction modalities within direct (virtual finger vs. controller touch) and distant (eye-gaze vs. head-gaze vs. controller ray-casting) selection on movement time and the application-defined Fitts Accuracy Index; (2) to estimate the associations of age, treated continuously, with psychomotor and cognitive-flexibility performance, and to test whether age-related slowing differs by modality (Age \ensuremath{\times} Mode); (3) to examine whether digital literacy is associated with performance or moderates the age association; and (4) to test the processing-speed account by examining the Age \ensuremath{\times} Part interaction on the TMT-VR, the incremental predictive value of a Fitts-derived speed score, and the share of the age effect on TMT-B statistically accounted for by TMT-A. In line with the motor-ageing literature, it was hypothesised that age would be associated with slower movement times across modalities, with possible differential slowing across techniques; in line with the processing-speed account, it was hypothesised that the age association with the TMT-VR would be speed-general and partially carried by individual differences in basic speed. The digital-literacy analyses were nondirectional individual-difference tests, and analyses of the application-defined Accuracy Index were exploratory because the index does not consistently indicate closer or less variable pointing. A post hoc exploratory extension additionally examined the complete exported and derived TMT-VR battery (\S{}2.4) to test whether the cross-domain association and the digital-literacy association are general or specific to temporal efficiency.

\section{Materials and Methods}\label{sec:2}

\subsection{Participants}\label{sec:2.1}

The final sample comprised 202 adults aged 19--90 years (\textit{M = 44.3, SD = 18.3}; 114 women, 88 men) with 6--24 years of formal education (\textit{M} = 15.5, \textit{SD} = 3.0) and digital-skills scores ranging widely across the sample (observed 12--59 of a possible 12--72; \textit{M} = 40.5, \textit{SD} = 9.5). All participants completed the five-mode Fitts battery, the TMT-VR, and the person-level questionnaires; data collection took place between March and August 2026 at the Psychology Network Lab (PsyNet Lab) of the American College of Greece, the Experimental Psychology Lab of the National and Kapodistrian University of Athens, the `Nestor' Psychogeriatric Association, and the Amaroussion Open Care Centre for Older Adults (KAPI). Age coverage was not uniform: decade counts were 70 (19--29), 18 (30--39), 28 (40--49), 40 (50--59), 20 (60--69), 23 (70--79), and 3 (80--90), so estimates in the ninth decade rest on very few observations. However, we also conducted a sensitivity analysis excluding participants over 80 years to verify that the sparse oldest band did not drive the primary estimates.

All participants had normal or corrected-to-normal vision and normal upper-limb mobility. As in previous TMT-VR studies \parencite{ref26,ref27}, exclusion criteria were assessed by self-report and comprised a history of neurological or severe motor disorders, medication that may affect cognitive or motor functioning, visual impairment interfering with the use of VR equipment, and a diagnosed history of epilepsy, vertigo, or nausea, or prior cybersickness. Participants provided written informed consent; the study was approved by the Research Ethics and Deontology Committee (\ensuremath{\mathrm{E}}.\ensuremath{\mathrm{H}}.\ensuremath{\Delta}.\ensuremath{\mathrm{E}}.) of the National and Kapodistrian University of Athens (protocol No 20579/5.3.2024; approval decision No 152/4.4.2024) and by the Institutional Review Board of The American College of Greece (expedited review, protocol \#202602564; approved 16 February 2026).

\subsection{XR apparatus, software, and tasks}\label{sec:2.2}

Both XR tasks (i.e., Fitts' law task and TMT-VR) were administered on the same standalone head-mounted display (HTC VIVE Focus Vision, professional edition; HTC Corporation, Taoyuan, Taiwan), which has two 2448 \ensuremath{\times} 2448-pixel LCD panels (one per eye), a 90 Hz refresh rate, a field of view of up to 120\textdegree{}, automatic interpupillary-distance adjustment, integrated eye tracking with 120 Hz binocular gaze output, and inside-out tracking of the headset and its two handheld controllers. The applications were developed in Unity 2022.3 LTS with the XR Interaction Toolkit and OpenXR.

\subsubsection{Fitts' law task}\label{sec:2.2.1}

The Fitts task required participants to select circular targets as quickly and accurately as possible under five interaction modes: eye-gaze, head-gaze, and controller ray-casting (distant selection, targets at 4.00 m), and virtual finger and controller direct touch (direct selection, targets at 0.50 m). In every mode, the target to be selected was highlighted in yellow and was selected by holding the pointer steadily on it for 0.5 s, without any button press, so that a discrete confirmation action could not displace the pointer at the moment of selection \parencite{ref8,ref45}; dwell time accumulated across re-entries, so leaving a target before confirmation did not reset the timer; on confirmation the target turned blue and a sound was played, and the next target was highlighted. Twelve targets, all visible simultaneously, were arranged in a circle on a vertical plane and selected in a star-like sequence. The task drew on the general ergonomic principles for physical input devices in ISO 9241-400:2007 and adapted the multidirectional pointing arrangement in Annex B of ISO/TS 9241-411:2012 to the five immersive-VR modes.

Targets varied factorially in width (small, large) and centre-to-centre distance (short, long), yielding four ID cells per mode and 20 cells per participant; ID (in bits) was computed as log2(\textit{D}/\textit{W} + 1) \parencite{ref6,ref7} from the target distance \textit{D} and width \textit{W}. In the direct session, the targets were 2 or 4 cm in diameter and 40 or 60 cm apart (ID = 3.46, 4.00, 4.39, and 4.95 bits); in the distant session, they were 12 or 24 cm in diameter and 160 or 240 cm apart, and ID was computed from the corresponding visual angles at 4.00 m (\textit{W} = 1.72\textdegree{} or 3.44\textdegree{}; \textit{D} = 22.6\textdegree{} or 33.4\textdegree{}; ID = 2.92, 3.42, 3.82, and 4.35 bits), as recorded by the application. Familiarisation trials with five targets preceded the recorded rounds; each cell comprised a warm-up round and a recorded round of 12 selections, and only the recorded round was analysed.

The application recorded movement time (MT), reaction time, total task time, and the application-defined Fitts Accuracy Index (AI), computed as AI = We \ensuremath{-} DC with We = 4.133 \ensuremath{\times} SD(DC), where DC is the distance of the pointer from the target centre sampled during the dwell period and We is the effective width \parencite{ref8}; by the application's definition, higher values denote better performance, and negative values are permissible. The index was used in an earlier VR Fitts study, in which it was analysed alongside movement time, which followed Fitts' law, and perceived difficulty \parencite{ref8}. Because greater variability in DC increases We, the index combines variability in distance from the target centre and centring in a single value and does not consistently indicate closer or less variable pointing; moreover, because it is computed in application units that scale with target size, its value in the present task was governed mainly by target width, with target distance contributing little once width was held constant (\S{}3.3). It was therefore treated as an exploratory composite (\S{}2.4). Movement time, total task time and the accuracy index were the dependent variables of the analyses reported below; reaction time, which is a small and near-constant component of total task time, is reported descriptively in \S{}3. Figure 1 illustrates the Fitts task design.

\begin{figure*}[!t]\centering
\includegraphics[width=\textwidth]{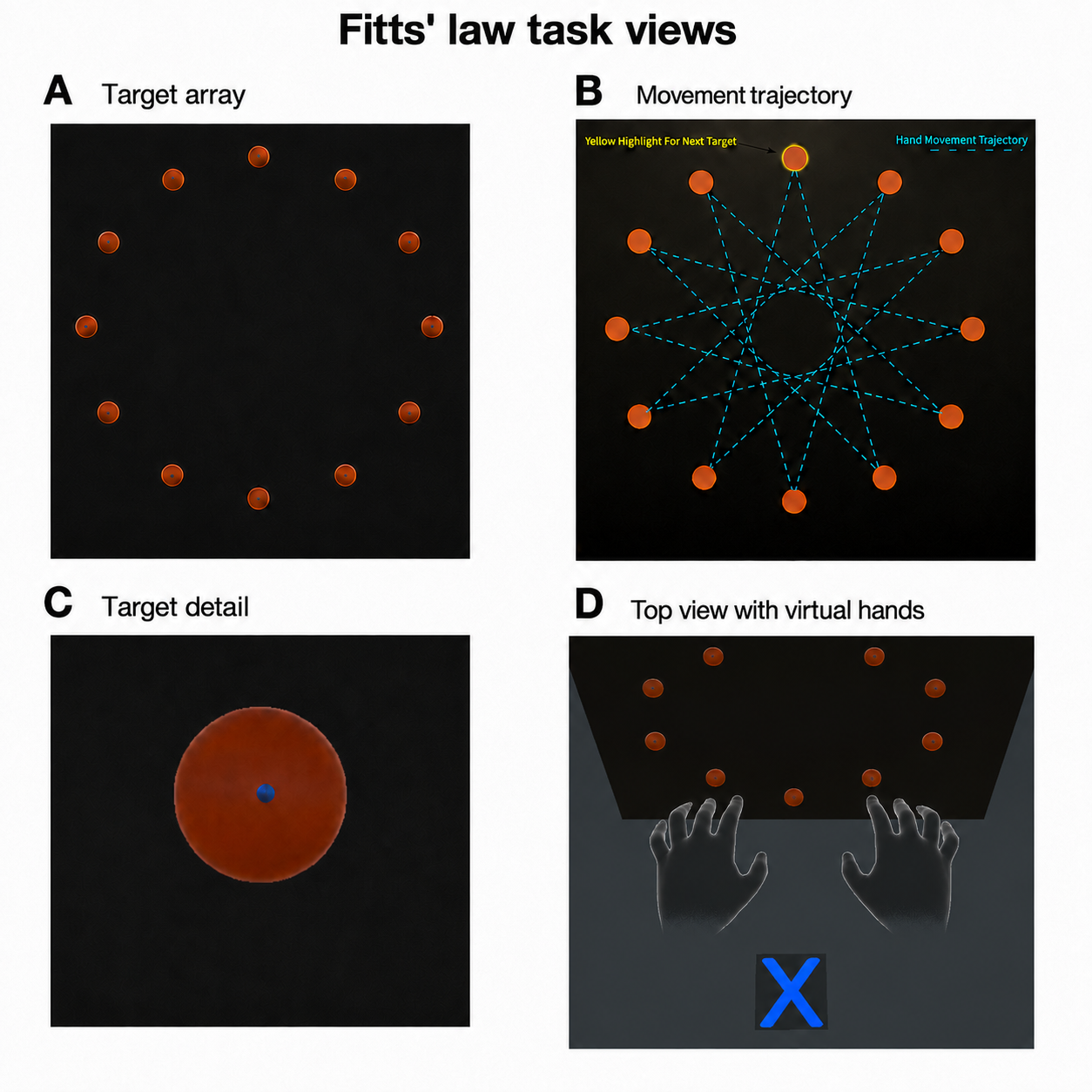}
\caption{Fitts' law task views. (A) Circular target array. (B) Movement trajectory between targets, with the next target highlighted in yellow. (C) Detail of an individual target. (D) Top view of the target array with the virtual hands visible (x is the position of the user on the floor). Panel B is adapted from the upper panel of Figure 2 in \textcite{ref8}. Panels are not shown at a common scale. An English-language \href{https://www.youtube.com/watch?v=I2VtC-1jV9k\&t=29s}{direct-mode demonstration} from the cited 2022 study (00:29--00:53) illustrates a task comparable to the present direct-mode condition.}\label{fig:1}
\end{figure*}

\subsubsection{Trail Making Test in VR (TMT-VR)}\label{sec:2.2.2}

The TMT-VR reproduces Parts A (numbers, 1--25) and B (alternating numbers and letters) in an immersive environment, with completion times recorded automatically by the application. All participants completed the TMT-VR with the head-gaze interaction mode on the same head-mounted display, seated (participants could stand briefly to locate a target). A target was selected by holding the head-gaze pointer on it for 0.5 s, without a button press, so that confirming a selection could not disturb the pointing itself \parencite{ref8,ref45}; the dwell restarted if the pointer left the target before confirmation; a correct selection was confirmed by one sound and a wrong selection was signalled by another, after which the participant continued, redirected by the examiner when necessary, until the correct target was selected, and each part ended only when all targets had been selected in the correct order. The target layout was rearranged between Parts A and B, and two practice trials preceded each part.

The examiner gave standardised instructions in Greek before the participant entered VR and during the practice trials, with additional clarification when needed, and the application also presented in-app instructions in Greek. Part-specific times were derived from the recorded total and difference scores (A = [Total \ensuremath{-} Difference]/2; B = [Total + Difference]/2). For each part the application also recorded the number of wrong-target selections and the mean selection distance---the Euclidean distance, in application units, from the registered ray-impact point to the target centre, averaged over correct selections only (smaller values denote selections closer to the target centre)---which were included in the exploratory extension described in \S{}2.4. Examples of the task displays are shown in Figure 2.

\begin{figure*}[!t]\centering
\includegraphics[width=.94\textwidth]{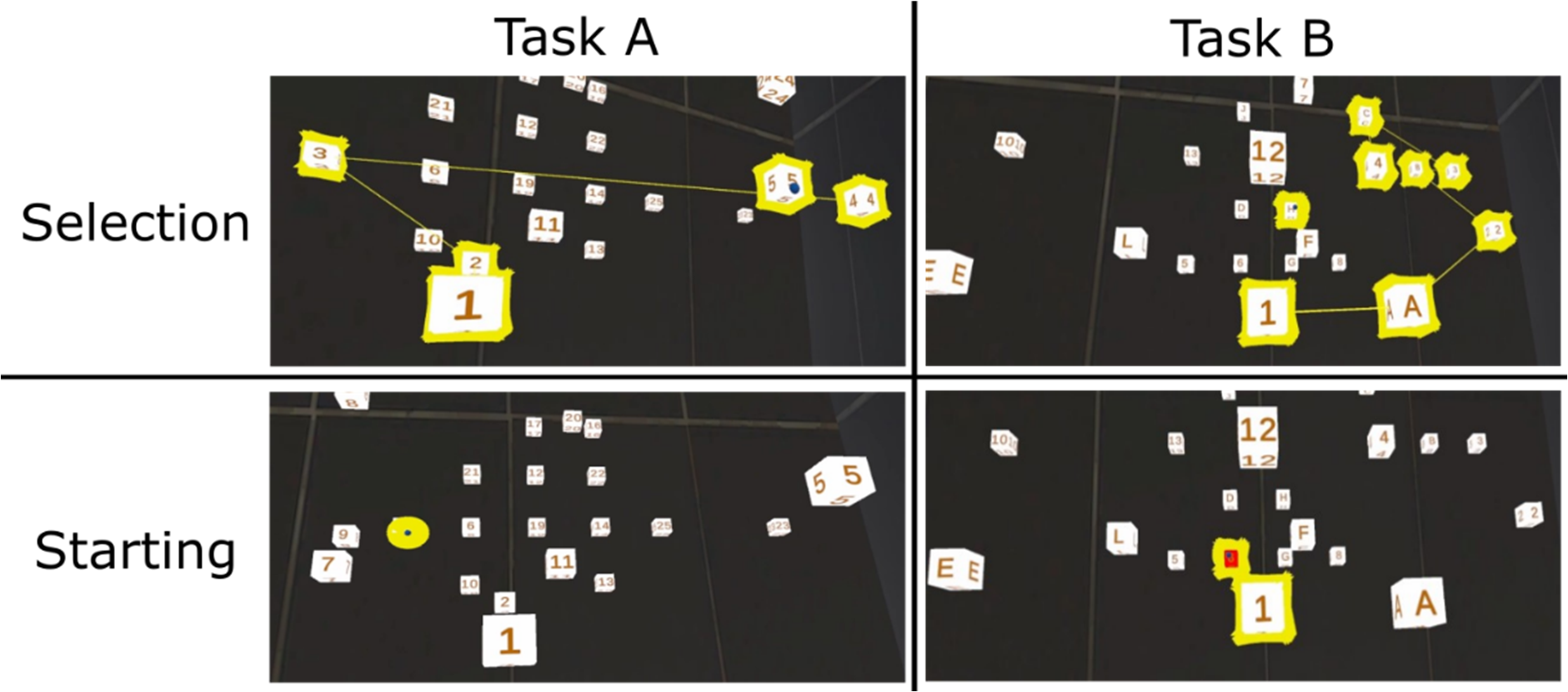}
\caption{TMT-VR task displays. Initial displays and target selections in Part A (numerical sequencing) and Part B (alternating numerical and alphabetical sequencing), reproduced from \textcite{ref27}, Figure 1, under the Creative Commons Attribution 4.0 licence. English-language demonstrations: \href{https://www.youtube.com/watch?v=npki7i4OnwY}{Task A video} and \href{https://www.youtube.com/watch?v=immvIkOyVuA}{Task B video}.}\label{fig:2}
\end{figure*}

\subsubsection{Digital literacy}\label{sec:2.2.3}

Digital literacy was measured with the Digital Skills Questionnaire (DSQ) used in previous VR research \parencite{ref27,ref46}. The DSQ comprises 12 items organised as six pairs (one frequency item and one proficiency item) across six device and application categories (computer, video games, smartphone, everyday smartphone applications, tablet, and VR), each rated on a six-point scale (frequency: 1 = less than once a month to 6 = daily; proficiency: 1 = no experience to 6 = professional). Total scores range from 12 to 72, with higher scores indicating greater digital skills (Cronbach's \ensuremath{\alpha} = .83 and McDonald's \ensuremath{\omega} = .85).

\subsection{Procedure}\label{sec:2.3}

Each participant attended a single session of approximately 45--50 min. After being informed about the study and providing written consent, participants completed the demographic questionnaire, the DSQ, and the Cybersickness in Virtual Reality Questionnaire (CSQ-VR) \parencite{ref47} on a tablet. The examiner then fitted and adjusted the headset, ran the eye-tracking calibration, and explained the interaction modes. Figure 3 shows the seated setup and examples of virtual-finger, controller ray-casting, and head-gaze selection. The dominant hand was registered by touching a virtual cube with the controller and was used for the finger and controller modes. The two VR tasks followed in an order counterbalanced across participants, with a short break between them; within the Fitts task, the order of the direct and distant sessions and the order of the interaction modes within each session were also counterbalanced, and the order of the four difficulty conditions within each mode block was varied systematically across participants. Automated audio instructions guided the tasks, and participants could pause between trials and tasks whenever they wished. The examiner monitored participants for signs of cybersickness throughout, and the session was to be stopped at the participant's request or at any sign of discomfort. After the VR tasks, participants completed the CSQ-VR a second time and were debriefed.

\begin{figure*}[!t]\centering
\includegraphics[width=.83\textwidth]{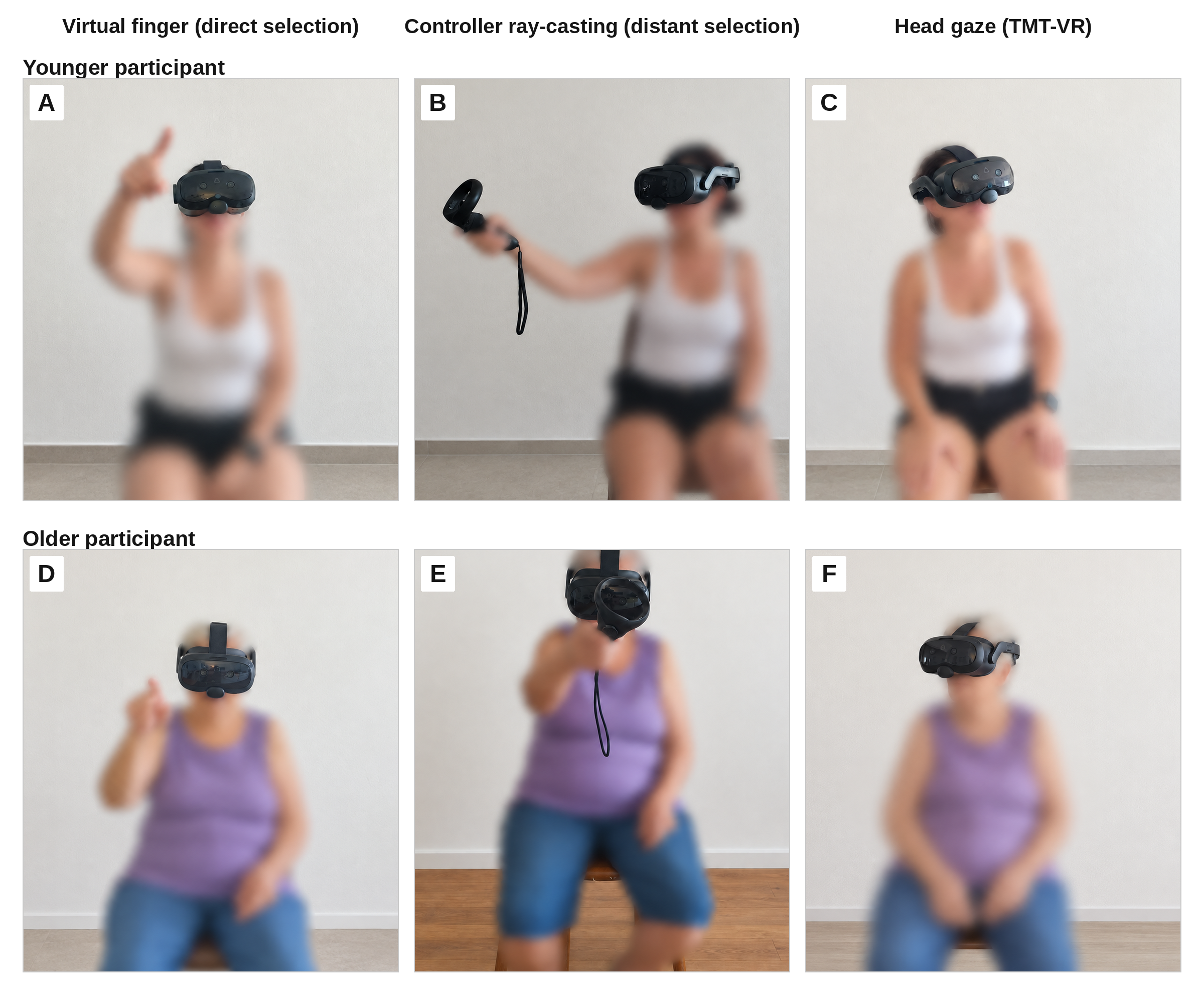}
\caption{Participant setup and examples of the interaction modes. A younger participant (A--C) and an older participant (D--F) seated and wearing the standalone HTC VIVE Focus Vision headset. (A, D) Virtual-finger selection, one of the two direct-selection modes of the Fitts' law task. (B, E) Controller ray-casting, one of the three distant-selection modes of the Fitts' law task. (C, F) Head-gaze selection in the TMT-VR, with the hands resting on the thighs. The photographs show the physical setup; the virtual displays are not shown. The backgrounds and the participants' faces and bodies were altered with AI-assisted image editing to protect the participants' identity (facial and bodily features, personal belongings); the headset and controller were not edited.}\label{fig:3}
\end{figure*}

\subsection{Statistical analyses}\label{sec:2.4}

\textit{General approach. }Analyses were conducted in R 4.4.1 \parencite{ref48} and RStudio \parencite{ref49} using reproducible scripts archived with the study materials; mixed models were fitted with lme4 \parencite{ref50} (Gaussian) and glmmTMB \parencite{ref51} (Gamma), marginal trends and contrasts were estimated with emmeans \parencite{ref52}, fit indices with performance \parencite{ref53}, and the exploratory factor model with lavaan \parencite{ref54}. Dependent variables were normalised with bestNormalize \parencite{ref55} (ordered-quantile transformation selected) and standardised. Gaussian linear mixed-effects models included participant random intercepts; Gamma mixed models with a log link provided raw-scale robustness checks. Age, digital literacy, education, and index of difficulty were entered as standardised continuous predictors; categorical factors used sum-to-zero contrasts. Added terms were evaluated by likelihood-ratio tests (LRTs), with AIC, BIC, and marginal/conditional R\textsuperscript{2} used to compare candidate models. Alpha was .05, two-tailed.

\textit{Fitts difficulty-cell analyses. }The 4,040 difficulty-cell observations (202 participants \ensuremath{\times} 5 modes \ensuremath{\times} 4 cells) were normalised before separation into the direct and distant sessions, and planned within-session contrasts used Bonferroni adjustment. Each session's chain added ID, Mode, Age, and digital literacy in turn and, in the distant session, Age \ensuremath{\times} Mode and Age \ensuremath{\times} digital literacy.

\textit{Five-mode person-level analysis. }The five-mode analysis used all 1,010 person-by-mode task times in one common transformation and one Mode \ensuremath{\times} Age model; per-mode age slopes and Tukey-adjusted pairwise contrasts were estimated with emmeans. Log-seconds and raw-seconds models tested scale sensitivity, and excluding participants over 80 tested leverage from the sparse oldest band. The application-defined Accuracy Index was analysed in the same way as an exploratory outcome.

\textit{TMT-VR and cross-domain analyses. }TMT-VR completion times were modelled at the part level with Part, Age, digital literacy, and Age \ensuremath{\times} Part. For the cross-domain analysis, participant random-intercept estimates from the difficulty-cell Fitts model of all five modes provided individual movement-time scores; the mean of the five standardised mode times (Cronbach's \ensuremath{\alpha} = .80) served as a complementary person-level composite. TMT-VR models tested the incremental contribution of these scores beyond age, and 2,000-resample percentile bootstrapping estimated the proportion of the age association with TMT-B statistically accounted for by TMT-A.

\textit{Exploratory full-battery extension.} After the planned analyses were complete, a post hoc exploratory extension examined the complete TMT-VR battery---the 26 exported and derived variables (per-part completion time, wrong-target errors, error rate, correctness accuracy, error-adjusted time, and mean selection distance; their totals across parts; the B \ensuremath{-} A differences; the B/A time ratio; and the normalised difference (B \ensuremath{-} A)/A; per part, error rate = errors/24, correctness accuracy = 24/(24 + errors), and adjusted time = time/accuracy)---as outcomes for the Fitts-derived predictors (the difficulty-cell movement-time score, the five-mode composite, and the five single-mode task times), adjusting for age, and for digital literacy, adjusting for age and education, with predictor \ensuremath{\times} Part interactions at the part level; every model used the same 202 participants.

Outcomes other than part-level and total wrong-target counts were normalised (ordered-quantile transform), standardised, and analysed with Gaussian linear mixed models (part level, participant random intercepts) or Gaussian linear models (person level). Part-level and total wrong-target counts were analysed with negative-binomial (nbinom2) models, with a participant random intercept at the part level; their coefficients are reported as log rate ratios. The signed B \ensuremath{-} A error difference, which can take negative values, was normalised and analysed with a Gaussian linear model. The resulting 208 likelihood-ratio tests (182 for the Fitts-derived predictors and 26 for digital literacy) were grouped into families by predictor family (the difficulty-cell score; the five-mode composite; the five single-mode task times pooled; digital literacy), outcome family (time, errors, adjusted time, selection distance, and the algebraic re-expressions error rate, correctness accuracy, and normalised difference), and test type (main effect or predictor \ensuremath{\times} Part), with Holm correction within family as the principal criterion and Benjamini--Hochberg values as a secondary descriptor; the algebraic re-expressions are deterministic functions of the time and error variables and were not counted as independent evidence. The extension is exploratory and was not preregistered; all 208 tests are listed in Supplementary Table S1.

\section{Results}\label{sec:3}

Table 1 presents descriptive statistics for the final sample.

Two further recorded measures are reported here for completeness. Reaction time, which was not modelled, averaged 0.34 s (\textit{SD} = 0.06) for eye-gaze, 0.45 s (\textit{SD} = 0.11) for head-gaze, 0.47 s (\textit{SD} = 0.12) for controller ray-casting, 0.65 s (\textit{SD} = 0.15) for controller direct touch, and 0.71 s (\textit{SD} = 0.15) for the virtual finger, and accounted for between 1.0\% and 2.8\% of the corresponding total task time. Cybersickness was minimal. On the CSQ-VR completed after the session, 60 of the 202 participants (29.7\%) rated every one of the six symptoms as absent (1 on a scale anchored 1 = absent to 7 = extreme); oculomotor symptoms were the most frequently reported (41.1\% and 50.5\% of participants across the two items), nausea (16.3\% and 32.7\%) and vestibular symptoms (22.3\% and 13.4\%) less so. The symptoms that were reported were mostly mild: 29 participants (14.4\%) rated any symptom at 4 or above and 11 (5.4\%) at 5 or above, and total scores ranged from 6 to 24 of a possible 6 to 42, with no participant above the scale midpoint and 179 (88.6\%) at 12 or below. No session was stopped for discomfort, and all 202 participants completed both VR tasks.

\begin{table*}[!t]
\caption{\textit{Descriptive Statistics for the Final Sample (N = 202)}}\label{tab:1}
\centering\small\setlength{\tabcolsep}{4pt}\renewcommand{\arraystretch}{1.18}
\begin{tabular}{L{0.52000\dimexpr\textwidth-24pt\relax}L{0.23000\dimexpr\textwidth-24pt\relax}L{0.25000\dimexpr\textwidth-24pt\relax}}
\toprule
\textbf{Measure} & \textbf{\textit{M} (\textit{SD})} & \textbf{Range} \\
\midrule
Age (years) & 44.3 (18.3) & 19--90 \\
Education (years) & 15.5 (3.0) & 6--24 \\
Digital skills (total) & 40.5 (9.5) & 12--59 \\
Eye-gaze task time (s) & 40.6 (21.4) & 20.5--142.9 \\
Head-gaze task time (s) & 29.4 (3.9) & 22.0--42.3 \\
Controller ray-cast task time (s) & 24.5 (4.4) & 17.9--57.3 \\
Controller direct task time (s) & 23.6 (5.2) & 17.1--64.5 \\
Virtual finger task time (s) & 26.0 (5.1) & 16.3--51.3 \\
TMT-VR Part A time (s) & 93.5 (48.1) & 43.2--334.5 \\
TMT-VR Part B time (s) & 111.8 (54.9) & 45.9--439.1 \\
\bottomrule\end{tabular}
\par\vspace{4pt}\begin{minipage}{\textwidth}\footnotesize \textit{Note.} TMT-VR = Trail Making Test in Virtual Reality. Task times are person-level means per interaction mode.\end{minipage}
\end{table*}

\subsection{Direct session (virtual finger vs. controller touch)}\label{sec:3.1}

Table 2 summarises the incremental model chains fitted to the difficulty-cell data of the direct session (202 participants \ensuremath{\times} 2 modes \ensuremath{\times} 4 cells = 1,616 observations). For MT, ID (\ensuremath{\chi}\textsuperscript{2}(1) = 350.21, \textit{p} < .001; \ensuremath{\beta} = 0.28), Mode (\ensuremath{\chi}\textsuperscript{2}(1) = 171.45, \textit{p} < .001), and Age (\ensuremath{\chi}\textsuperscript{2}(1) = 92.04, p < .001; \ensuremath{\beta} = 0.32) each improved fit, whereas DSQ did not (\ensuremath{\chi}\textsuperscript{2}(1) = 0.03, p = .857); the age-plus-design model was retained by AIC and BIC. Selection with the bare virtual finger was slower than controller touch (\ensuremath{\Delta} = +0.33 \textit{SD}; \textit{t}(1416) = 13.49, \textit{p} < .001, \textit{d} = 0.67). The accuracy index showed the same structure: ID (\ensuremath{\chi}\textsuperscript{2}(1) = 152.63, \textit{p} < .001), Mode (\ensuremath{\chi}\textsuperscript{2}(1) = 25.31, \textit{p} < .001), and Age (\ensuremath{\chi}\textsuperscript{2}(1) = 22.86, p < .001; \ensuremath{\beta} = 0.07) improved fit; DSQ did not (\ensuremath{\chi}\textsuperscript{2}(1) = 0.03, p = .860).

\begin{table*}[!t]
\caption{\textit{Incremental Model Chains, Direct Session (Difficulty-Cell Level, N = 202)}}\label{tab:2}
\centering\small\setlength{\tabcolsep}{4pt}\renewcommand{\arraystretch}{1.18}
\begin{tabular}{L{0.24000\dimexpr\textwidth-48pt\relax}L{0.28000\dimexpr\textwidth-48pt\relax}L{0.14000\dimexpr\textwidth-48pt\relax}L{0.06000\dimexpr\textwidth-48pt\relax}L{0.14000\dimexpr\textwidth-48pt\relax}L{0.14000\dimexpr\textwidth-48pt\relax}}
\toprule
\textbf{Outcome} & \textbf{Step} & \textbf{\ensuremath{\chi}\textsuperscript{2}} & \textbf{\textit{df}} & \textbf{\textit{p}} & \textbf{\ensuremath{\beta}} \\
\midrule
Movement time & + ID & 350.21 & 1 & < .001 & 0.28 \\
Movement time & + Mode & 171.45 & 1 & < .001 & --- \\
Movement time & + Age & 92.04 & 1 & < .001 & 0.32 \\
Movement time & + Digital skills & 0.03 & 1 & .857 & --- \\
Accuracy index & + ID & 152.63 & 1 & < .001 & --- \\
Accuracy index & + Mode & 25.31 & 1 & < .001 & --- \\
Accuracy index & + Age & 22.86 & 1 & < .001 & 0.07 \\
Accuracy index & + Digital skills & 0.03 & 1 & .860 & --- \\
\bottomrule\end{tabular}
\par\vspace{4pt}\begin{minipage}{\textwidth}\footnotesize \textit{Note.} Likelihood-ratio tests for each added term; \ensuremath{\beta} = standardised coefficient at entry. ID = index of difficulty.\end{minipage}
\end{table*}

\subsection{Distant session (eye-gaze vs. head-gaze vs. ray-casting)}\label{sec:3.2}

For MT (Table 3; 202 participants \ensuremath{\times} 3 modes \ensuremath{\times} 4 cells = 2,424 observations), ID (\ensuremath{\chi}\textsuperscript{2}(1) = 1164.66, \textit{p} < .001; \ensuremath{\beta} = 0.74), Mode (\ensuremath{\chi}\textsuperscript{2}(2) = 740.58, \textit{p} < .001), and Age (\ensuremath{\chi}\textsuperscript{2}(1) = 45.80, p < .001; \ensuremath{\beta} = 0.20) improved fit; DSQ did not (\ensuremath{\chi}\textsuperscript{2}(1) = 3.66, p = .056). The Age \ensuremath{\times} Mode step improved fit (\ensuremath{\chi}\textsuperscript{2}(2) = 10.22, p = .006), whereas Age \ensuremath{\times} DSQ did not (\ensuremath{\chi}\textsuperscript{2}(1) = 0.02, p = .875). The interaction model was retained by LRT and AIC (BIC favoured the main-effects model; both are reported). Planned contrasts indicated that gaze-based selection was slower than ray-casting overall (+0.81 \textit{SD}; \textit{t}(2225) = 27.64, \textit{p} < .001) and that eye-gaze was slower than head-gaze (+0.36 \textit{SD}; \textit{t}(2225) = 10.67, \textit{p} < .001). Within this session, age-related slowing was steepest for ray-casting (\ensuremath{\beta} = 0.29, 95\% CI [0.22, 0.36]) and shallower, but clearly present, for head-gaze (\ensuremath{\beta} = 0.20 [0.13, 0.28]) and eye-gaze (\ensuremath{\beta} = 0.19 [0.12, 0.26]). For the accuracy index, only ID (\ensuremath{\chi}\textsuperscript{2}(1) = 83.93, \textit{p} < .001) and Mode (\ensuremath{\chi}\textsuperscript{2}(2) = 1664.88, \textit{p} < .001) improved fit; Age (\ensuremath{\chi}\textsuperscript{2}(1) = 2.53, p = .112), DSQ (\ensuremath{\chi}\textsuperscript{2}(1) = 0.05, \textit{p} = .822), and Age \ensuremath{\times} Mode (\ensuremath{\chi}\textsuperscript{2}(2) = 0.34, p = .845) did not.

\begin{table*}[!t]
\caption{\textit{Incremental Model Chains, Distant Session (Difficulty-Cell Level, N = 202)}}\label{tab:3}
\centering\small\setlength{\tabcolsep}{4pt}\renewcommand{\arraystretch}{1.18}
\begin{tabular}{L{0.24000\dimexpr\textwidth-48pt\relax}L{0.28000\dimexpr\textwidth-48pt\relax}L{0.14000\dimexpr\textwidth-48pt\relax}L{0.06000\dimexpr\textwidth-48pt\relax}L{0.14000\dimexpr\textwidth-48pt\relax}L{0.14000\dimexpr\textwidth-48pt\relax}}
\toprule
\textbf{Outcome} & \textbf{Step} & \textbf{\ensuremath{\chi}\textsuperscript{2}} & \textbf{\textit{df}} & \textbf{\textit{p}} & \textbf{\ensuremath{\beta}} \\
\midrule
Movement time & + ID & 1164.66 & 1 & < .001 & 0.74 \\
Movement time & + Mode & 740.58 & 2 & < .001 & --- \\
Movement time & + Age & 45.80 & 1 & < .001 & 0.20 \\
Movement time & + Digital skills & 3.66 & 1 & .056 & --- \\
Movement time & + Age \ensuremath{\times} Mode & 10.22 & 2 & .006 & --- \\
Movement time & + Age \ensuremath{\times} Digital skills & 0.02 & 1 & .875 & --- \\
Accuracy index & + ID & 83.93 & 1 & < .001 & --- \\
Accuracy index & + Mode & 1664.88 & 2 & < .001 & --- \\
Accuracy index & + Age & 2.53 & 1 & .112 & --- \\
Accuracy index & + Digital skills & 0.05 & 1 & .822 & --- \\
Accuracy index & + Age \ensuremath{\times} Mode & 0.34 & 2 & .845 & --- \\
\bottomrule\end{tabular}
\par\vspace{4pt}\begin{minipage}{\textwidth}\footnotesize \textit{Note.} Likelihood-ratio tests for each added term; \ensuremath{\beta} = standardised coefficient at entry. ID = index of difficulty.\end{minipage}
\end{table*}

\subsection{Five-mode person-level analysis}\label{sec:3.3}

Person-level task times for the five modes (202 participants \ensuremath{\times} 5 modes = 1,010 observations; normalised with a single common transform and standardised; Figure 4) were modelled with mode fixed effects and participant random intercepts (Table 4). Age improved fit strongly (\ensuremath{\chi}\textsuperscript{2}(1) = 104.48, \textit{p} < .001; \ensuremath{\beta} = 0.36; marginal R\textsuperscript{2} = .50), whereas digital literacy did not (\ensuremath{\chi}\textsuperscript{2}(1) = 1.58, p = .209). The Age \ensuremath{\times} Mode step was highly significant (\ensuremath{\chi}\textsuperscript{2}(4) = 70.83, \textit{p} < .001). Per-mode age slopes estimated from the common interaction model (estimated marginal trends; Kenward--Roger) were: controller direct touch \ensuremath{\beta} = 0.67 (95\% CI [0.58, 0.77]), controller ray-casting \ensuremath{\beta} = 0.36 [0.26, 0.45], virtual finger \ensuremath{\beta} = 0.30 [0.20, 0.39], eye-gaze \ensuremath{\beta} = 0.25 [0.15, 0.34], and head-gaze \ensuremath{\beta} = 0.24 [0.15, 0.34] (Figure 5). Tukey-adjusted pairwise slope contrasts showed that the controller direct slope exceeded every other mode's slope (all \textit{p} < .001), whereas no other pair of slopes differed reliably (smallest adjusted p = .311). The interaction was therefore driven by controller direct touch.

Because standardised slopes depend on the outcome scale, two sensitivity analyses were run. On log seconds the interaction persisted (\ensuremath{\chi}\textsuperscript{2}(4) = 16.81, p = .002), with estimated slowing per decade of 6.9\% [5.1, 8.7] for controller direct, 5.0\% [3.3, 6.8] for eye-gaze, 3.8\% [2.2, 5.6] for ray-casting, 3.3\% [1.6, 5.0] for head-gaze, and 2.8\% [1.2, 4.5] for the virtual finger; on raw seconds the interaction was not clear (\ensuremath{\chi}\textsuperscript{2}(4) = 7.15, p = .128). The distinct direct-touch gradient was robust to excluding the three participants over 80 (\ensuremath{\chi}\textsuperscript{2}(4) = 65.77, p < .001; direct slope \ensuremath{\beta} = 0.66). To translate the relative gradients into absolute terms, geometric-mean task times per mode were predicted from the log-seconds interaction model at representative ages: controller direct touch was fastest at age 30 (21.1 s [20.3, 21.9], against 23.0--34.3 s for the other modes) and remained comparable to the fastest modes at 70 (27.5 s [26.1, 29.0], against 26.7 s ray-cast and 27.5 s virtual finger) and 85 (30.4 s [28.2, 32.7], against 28.3--28.6 s), with eye-gaze slowest at every age. Tukey-adjusted contrasts at fixed ages showed direct touch significantly faster than every other mode at age 30, faster than head-gaze, eye-gaze, and the virtual finger at 50, faster than head- and eye-gaze at 70, and distinguishable only from eye-gaze at 85; its absolute advantage eroded with age but never reversed significantly. The difficulty-cell analyses (\S{}3.2) likewise showed mode-dependent age slopes within the distant session.

\begin{table*}[!t]
\caption{\textit{Person-Level Five-Mode Chain and Common-Model Age Slopes (N = 202)}}\label{tab:4}
\centering\small\setlength{\tabcolsep}{4pt}\renewcommand{\arraystretch}{1.18}
\begin{tabular}{L{0.36000\dimexpr\textwidth-32pt\relax}L{0.23000\dimexpr\textwidth-32pt\relax}L{0.08000\dimexpr\textwidth-32pt\relax}L{0.33000\dimexpr\textwidth-32pt\relax}}
\toprule
\textbf{Step / Mode} & \textbf{\ensuremath{\chi}\textsuperscript{2} or \ensuremath{\beta}} & \textbf{\textit{df}} & \textbf{\textit{p} / 95\% CI} \\
\midrule
+ Age & \ensuremath{\chi}\textsuperscript{2} = 104.48 & 1 & < .001 (\ensuremath{\beta} = 0.36) \\
+ Digital skills & \ensuremath{\chi}\textsuperscript{2} = 1.58 & 1 & .209 \\
+ Age \ensuremath{\times} Mode & \ensuremath{\chi}\textsuperscript{2} = 70.83 & 4 & < .001 \\
Controller direct & \ensuremath{\beta} = 0.67 & --- & [0.58, 0.77] \\
Controller ray-cast & \ensuremath{\beta} = 0.36 & --- & [0.26, 0.45] \\
Virtual finger & \ensuremath{\beta} = 0.30 & --- & [0.20, 0.39] \\
Eye-gaze & \ensuremath{\beta} = 0.25 & --- & [0.15, 0.34] \\
Head-gaze & \ensuremath{\beta} = 0.24 & --- & [0.15, 0.34] \\
\bottomrule\end{tabular}
\par\vspace{4pt}\begin{minipage}{\textwidth}\footnotesize \textit{Note.} Upper block: likelihood-ratio steps over the mode-plus-random-intercept baseline. Lower block: per-mode age slopes estimated from the single common Mode \ensuremath{\times} Age model (estimated marginal trends, Kenward--Roger 95\% confidence intervals). Tukey-adjusted contrasts: controller direct differs from every other mode (\textit{p} < .001); no other pair differs (smallest adjusted p = .311).\end{minipage}
\end{table*}

\begin{figure*}[!t]\centering
\includegraphics[width=.88\textwidth]{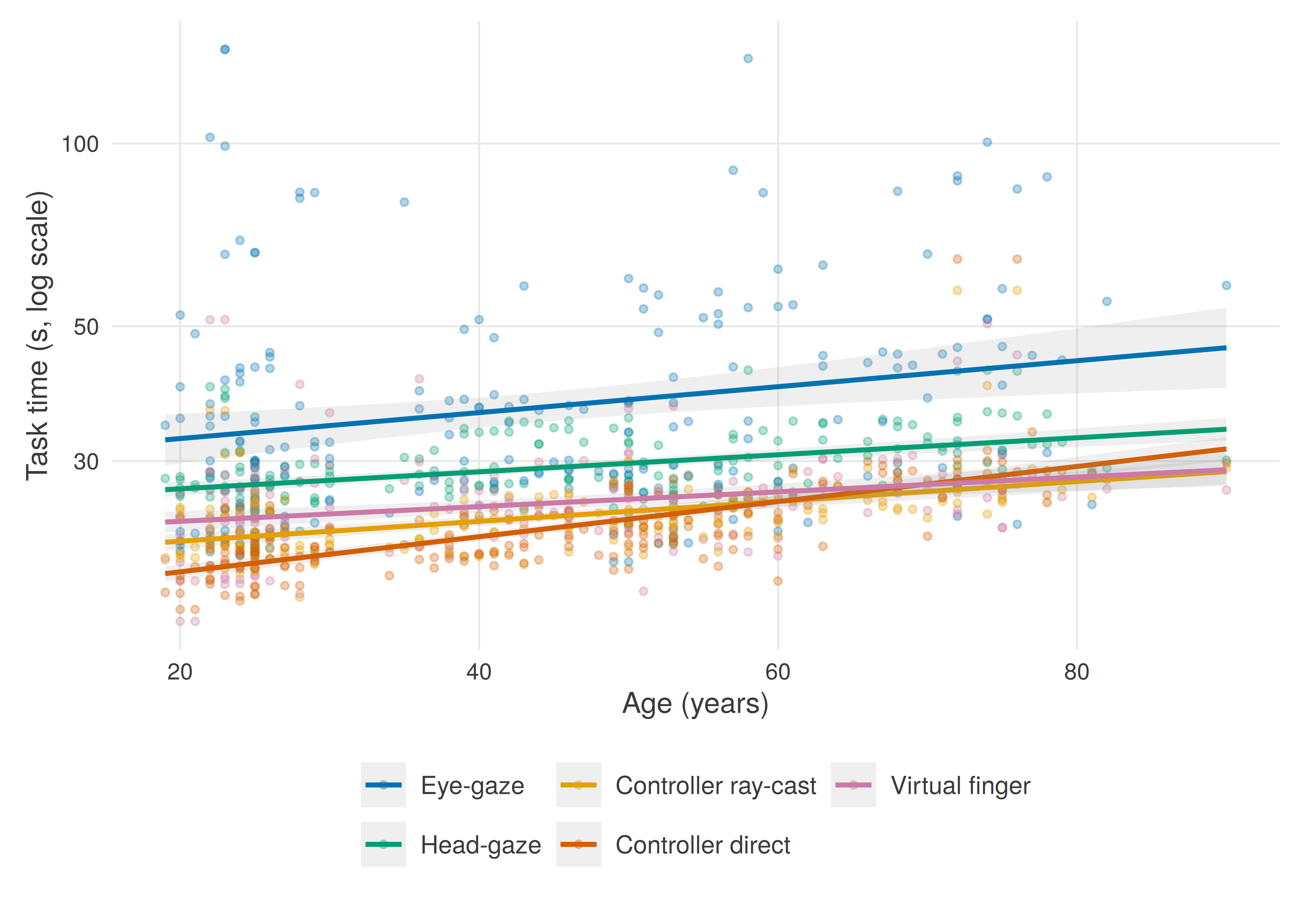}
\caption{\textit{Task Time by Age and Interaction Mode}. \textit{Note.} Person-level task time (\textit{N} = 202; logarithmic scale). Points are participant-by-mode means; lines are linear fits with 95\% confidence bands. Model-based estimates appear in the text and Table 4. Direct selection: controller direct touch and virtual finger; distant selection: eye-gaze, head-gaze, and controller ray-casting.}\label{fig:4}
\end{figure*}

\begin{figure*}[!t]\centering
\includegraphics[width=.88\textwidth]{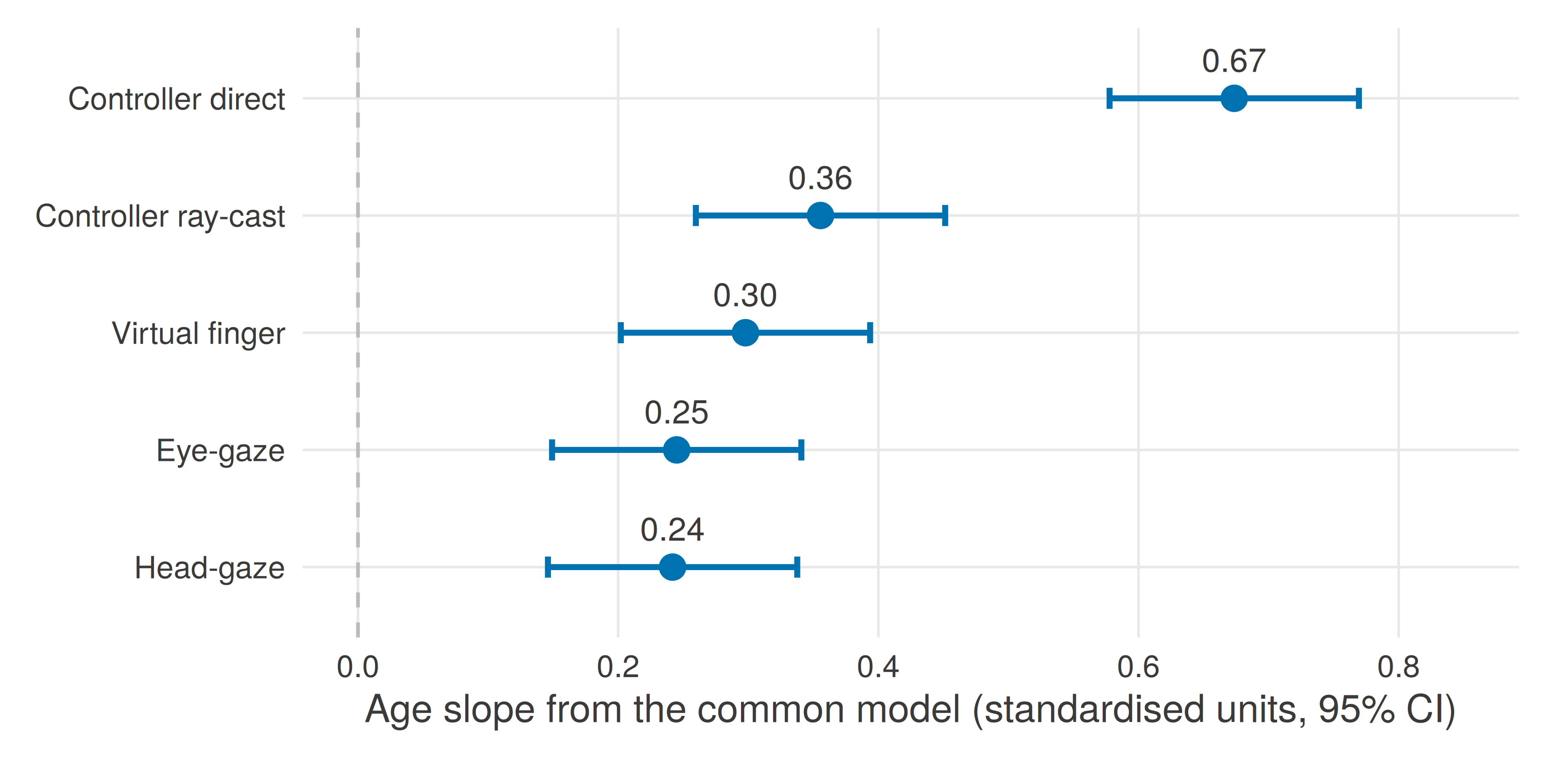}
\caption{\textit{Common-Model Age Slopes per Interaction Mode}. \textit{Note.} Age slopes (with Kenward--Roger 95\% confidence intervals) of normalised person-level task time, estimated from the single common Mode \ensuremath{\times} Age mixed model (\textit{N} = 202). The dashed line marks a null age effect. Only the controller direct slope differs reliably from the others (Tukey-adjusted \textit{p} < .001). Direct selection: controller direct touch and virtual finger; distant selection: eye-gaze, head-gaze, and controller ray-casting.}\label{fig:5}
\end{figure*}

The application-defined Fitts Accuracy Index (AI = We \ensuremath{-} DC; \S{}2.2.1) showed no overall age association (\ensuremath{\beta} = 0.00, 95\% CI [\ensuremath{-}0.04, 0.04]; \ensuremath{\chi}\textsuperscript{2}(1) = 0.00, p = .982) but a clear Age \ensuremath{\times} Mode interaction (\ensuremath{\chi}\textsuperscript{2}(4) = 52.81, p < .001). Common-model age slopes were: head-gaze \ensuremath{-}0.12 [\ensuremath{-}0.20, \ensuremath{-}0.04], controller ray-cast \ensuremath{-}0.11 [\ensuremath{-}0.19, \ensuremath{-}0.04], eye-gaze \ensuremath{-}0.02 [\ensuremath{-}0.10, 0.06], virtual finger +0.03 [\ensuremath{-}0.04, 0.11], and controller direct +0.22 [0.14, 0.30]. In Tukey-adjusted contrasts the controller-direct slope differed from every other mode (e.g., controller ray-cast \ensuremath{-} controller direct = \ensuremath{-}0.334, \textit{p} < .001), and the only other detected contrasts were head-gaze versus virtual finger (\ensuremath{-}0.155, p = .032) and virtual finger versus ray-casting (0.148, p = .048). Across the 4,040 difficulty cells, mode and target width together accounted for 92.5\% of the variance in the index and participant identity for a further 1.2\%; within a difficulty cell, the index varied between participants with a coefficient of variation of 1.2--2.1\% for head-gaze and ray-casting, so the slopes above describe small absolute shifts. Age was thus associated with lower index values for head-gaze and ray-casting but with higher values for controller direct touch, the mode with the steepest time gradient. Because the index can increase when variability in distance from the target centre increases, this pattern is interpreted cautiously (\S{}4.2).

\subsection{TMT-VR}\label{sec:3.4}

Part (Part B slower), Age (\ensuremath{\chi}\textsuperscript{2}(1) = 168.13, p < .001; \ensuremath{\beta} = 0.60), and DSQ (\ensuremath{\chi}\textsuperscript{2}(1) = 5.56, p = .018; \ensuremath{\beta} = \ensuremath{-}0.11, higher digital literacy associated with faster completion) improved fit. The Age \ensuremath{\times} Part step did not improve fit (\ensuremath{\chi}\textsuperscript{2}(1) = 0.04, p = .835; interaction \ensuremath{\beta} = 0.01, 95\% CI [\ensuremath{-}0.05, 0.06]), indicating similar age gradients for Parts A and B in this sample.

\subsection{Fitts-derived speed scores and cross-domain association}\label{sec:3.5}

The by-participant random slope for ID did not improve fit over a random-intercept model (\ensuremath{\chi}\textsuperscript{2}(2) = 2.21, p = .331), so the extracted score captured overall psychomotor speed. An exploratory one-factor model of the five normalised mode times fit poorly (robust CFI = .911, RMSEA = .176), precluding a latent-factor interpretation. Table 5 summarises the cross-domain results. In the chain that carries the difficulty-cell Fitts score, Age improved part-level TMT-VR model fit (\ensuremath{\chi}\textsuperscript{2}(1) = 168.14, p < .001; \ensuremath{\beta} = 0.66). This coefficient is estimated before the Fitts score is entered and without adjustment for digital literacy, and is therefore larger than the \ensuremath{\beta} = 0.60 of the digital-literacy-adjusted TMT-VR chain in Table 5. The Fitts movement-time score improved fit beyond age (\ensuremath{\chi}\textsuperscript{2}(1) = 13.20, p < .001; \ensuremath{\beta} = 0.18; marginal R\textsuperscript{2} .489 \ensuremath{\rightarrow} .510); the score did not interact with Part (p = .663). Adding the score attenuated the age coefficient by 15.7\% (\ensuremath{\beta} 0.66 \ensuremath{\rightarrow} 0.56), although age remained significant (\ensuremath{\chi}\textsuperscript{2}(1) = 101.90, p < .001). The person-level five-mode speed composite likewise improved the corresponding part-level model (\ensuremath{\chi}\textsuperscript{2}(1) = 10.00, p = .002; \ensuremath{\beta} = 0.17) and attenuated the age coefficient by 15.9\% (0.660 \ensuremath{\rightarrow} 0.555). Within the TMT-VR, TMT-A performance statistically accounted for 29.4\% of the age association with TMT-B (indirect association = 0.203, bootstrap 95\% CI [0.100, 0.306]; direct association 70.6\%, p < .001). A Gamma (log-link) robustness model reproduced the score association (b = 0.07, p < .001).

\begin{table*}[!t]
\caption{\textit{Cross-Domain Chain and Variance-Decomposition Summary (N = 202)}}\label{tab:5}
\centering\small\setlength{\tabcolsep}{4pt}\renewcommand{\arraystretch}{1.18}
\begin{tabular}{L{0.47000\dimexpr\textwidth-24pt\relax}L{0.28000\dimexpr\textwidth-24pt\relax}L{0.25000\dimexpr\textwidth-24pt\relax}}
\toprule
\textbf{Analysis} & \textbf{Estimate} & \textbf{\textit{p} / 95\% CI} \\
\midrule
TMT-VR chain: + Age & \ensuremath{\chi}\textsuperscript{2}(1) = 168.13, \ensuremath{\beta} = 0.60 & < .001 \\
TMT-VR chain: + Digital skills & \ensuremath{\chi}\textsuperscript{2}(1) = 5.56, \ensuremath{\beta} = \ensuremath{-}0.11 & .018 \\
TMT-VR chain: + Age \ensuremath{\times} Part & \ensuremath{\chi}\textsuperscript{2}(1) = 0.04, \ensuremath{\beta} = 0.01 [\ensuremath{-}0.05, 0.06] & .835 \\
Fitts movement-time score beyond age (difficulty-cell model) & \ensuremath{\chi}\textsuperscript{2}(1) = 13.20, \ensuremath{\beta} = 0.18 & < .001; age attenuation 15.7\% \\
Five-mode composite beyond age (part-level model) & \ensuremath{\chi}\textsuperscript{2}(1) = 10.00, \ensuremath{\beta} = 0.17 & .002; age attenuation 15.9\% \\
Age \ensuremath{\rightarrow} Part A \ensuremath{\rightarrow} Part B indirect association & 0.203 (29.4\% of total) & [0.100, 0.306] \\
\bottomrule\end{tabular}
\par\vspace{4pt}\begin{minipage}{\textwidth}\footnotesize \textit{Note.} The difficulty-cell score is the participant random-intercept estimate from the difficulty-cell Fitts model of all five modes (direct and distant sessions pooled); the five-mode composite is the mean of the five standardised mode times (\ensuremath{\alpha} = .80), tested in the same part-level mixed model as the age effect. The indirect association used 2,000 bootstrap resamples. All estimates are cross-sectional variance decompositions.\end{minipage}
\end{table*}

\subsection{Exploratory Analyses: specificity of the cross-domain association (post hoc)}\label{sec:3.6}

Across the 208 exploratory tests, fifteen reached Holm-adjusted p < .05, all in the completion-time and error-adjusted-time families (Table 6; Supplementary Table S1). The difficulty-cell Fitts movement-time score was associated, beyond age, with part-level completion time (b = 0.18 [0.09, 0.28], \ensuremath{\chi}\textsuperscript{2}(1) = 13.80, Holm p < .001), part-level error-adjusted time (b = 0.16 [0.06, 0.25], \ensuremath{\chi}\textsuperscript{2}(1) = 10.28, Holm p = .004), total time (b = 0.21 [0.10, 0.32], \ensuremath{\chi}\textsuperscript{2}(1) = 13.07, Holm p < .001), and total error-adjusted time (b = 0.18 [0.07, 0.29], \ensuremath{\chi}\textsuperscript{2}(1) = 9.37, Holm p = .004)---four views of one temporal-efficiency association rather than four independent findings. The five-mode composite showed the same four associations (Holm p = .005 to .015), and among the five single-mode task times head-gaze (three tests, Holm p = .025 to .038) and controller direct touch (two tests, Holm p = .021 and .034) were associated with part-level completion time and error-adjusted time. Digital literacy, beyond age and education, was associated with part-level error-adjusted time (b = \ensuremath{-}0.13 [\ensuremath{-}0.22, \ensuremath{-}0.04], \ensuremath{\chi}\textsuperscript{2}(1) = 7.23, Holm p = .022) and total error-adjusted time (b = \ensuremath{-}0.15 [\ensuremath{-}0.26, \ensuremath{-}0.04], \ensuremath{\chi}\textsuperscript{2}(1) = 7.19, Holm p = .022); its part-level and total completion-time associations fell just short of the criterion (Holm p = .052 for both).

No test in the wrong-target error, selection-distance, B \ensuremath{-} A difference, B/A ratio, or predictor \ensuremath{\times} Part families survived Holm correction for any predictor (smallest Holm-adjusted p = .069, for the virtual-finger \ensuremath{\times} Part term on part-level errors; .313 for the difficulty-cell score and .079 for digital literacy), and the algebraic re-expressions reproduced the null pattern of their parent variables. Eye-gaze, virtual-finger, and ray-casting task times were not associated with any TMT-VR outcome after correction (smallest Holm-adjusted p = .069). The Fitts-derived associations were therefore specific to the completion-time and error-adjusted-time indices, and the digital-literacy association to error-adjusted time.

\begin{table*}[!t]
\caption{\textit{Exploratory Full-Battery Extension: Temporal-Efficiency Associations of the Fitts-Derived Scores and Digital Literacy with the TMT-VR Battery}}\label{tab:6}
\centering\small\setlength{\tabcolsep}{4pt}\renewcommand{\arraystretch}{1.18}
\begin{tabular}{L{0.22000\dimexpr\textwidth-48pt\relax}L{0.26000\dimexpr\textwidth-48pt\relax}L{0.07500\dimexpr\textwidth-48pt\relax}L{0.08500\dimexpr\textwidth-48pt\relax}L{0.08500\dimexpr\textwidth-48pt\relax}L{0.27500\dimexpr\textwidth-48pt\relax}}
\toprule
\textbf{Predictor} & \textbf{Outcome} & \textbf{\ensuremath{\chi}\textsuperscript{2}(1)} & \textbf{Raw \textit{p}} & \textbf{Holm \textit{p}} & \textbf{\textit{b} [95\% CI]} \\
\midrule
Difficulty-cell Fitts score & Part-level completion time & 13.80 & < .001 & < .001 & 0.18 [0.09, 0.28] \\
Difficulty-cell Fitts score & Part-level error-adjusted time & 10.28 & .001 & .004 & 0.16 [0.06, 0.25] \\
Difficulty-cell Fitts score & Total time & 13.07 & < .001 & < .001 & 0.21 [0.10, 0.32] \\
Difficulty-cell Fitts score & Total error-adjusted time & 9.37 & .002 & .004 & 0.18 [0.07, 0.29] \\
Five-mode composite & Part-level completion time & 10.32 & .001 & .005 & 0.17 [0.07, 0.27] \\
Five-mode composite & Part-level error-adjusted time & 8.39 & .004 & .011 & 0.15 [0.05, 0.25] \\
Five-mode composite & Total time & 9.76 & .002 & .005 & 0.19 [0.07, 0.31] \\
Five-mode composite & Total error-adjusted time & 7.17 & .007 & .015 & 0.17 [0.04, 0.29] \\
Head-gaze task time & Part-level completion time & 10.29 & .001 & .025 & 0.15 [0.06, 0.23] \\
Head-gaze task time & Total time & 10.22 & .001 & .025 & 0.17 [0.07, 0.28] \\
Head-gaze task time & Part-level error-adjusted time & 8.97 & .003 & .038 & 0.14 [0.05, 0.22] \\
Controller direct task time & Part-level completion time & 10.78 & .001 & .021 & 0.20 [0.08, 0.32] \\
Controller direct task time & Part-level error-adjusted time & 9.32 & .002 & .034 & 0.19 [0.07, 0.31] \\
Digital literacy & Part-level error-adjusted time & 7.23 & .007 & .022 & \ensuremath{-}0.13 [\ensuremath{-}0.22, \ensuremath{-}0.04] \\
Digital literacy & Total error-adjusted time & 7.19 & .007 & .022 & \ensuremath{-}0.15 [\ensuremath{-}0.26, \ensuremath{-}0.04] \\
\bottomrule\end{tabular}
\par\vspace{4pt}\begin{minipage}{\textwidth}\footnotesize \textit{Note.} Post hoc exploratory analyses of the complete TMT-VR battery (26 variables; 208 tests; all in the same 202 participants); outcomes normalised (ordered-quantile transform) and standardised; Fitts-derived predictors adjusted for age, digital literacy for age and education; participant random intercepts at the part level. \textit{b} = standardised coefficient (\textit{SD} per predictor \textit{SD}); a positive \textit{b} denotes slower or less efficient TMT-VR performance with a slower Fitts score, a negative b denotes faster performance with higher digital literacy. Holm correction within family (predictor family \ensuremath{\times} outcome family \ensuremath{\times} test type; the five single-mode task times form one pooled family). The fifteen rows listed are the only Holm-significant tests. No error, selection-distance, difference, ratio, or predictor \ensuremath{\times} Part test survived correction for any predictor; all 208 tests are listed in Supplementary Table S1.\end{minipage}
\end{table*}

\section{Discussion}\label{sec:4}

The present study had four aims: to compare interaction modalities within direct and distant selection; to estimate the associations of age with psychomotor and cognitive-flexibility performance and to test whether age-related slowing differs by modality; to examine the role of digital literacy; and to test the processing-speed account of the association between the two tasks. Overall, four findings emerged. First, interaction modality substantially shaped pointing performance in both sessions. Second, age was associated with slower performance in every mode, and this slowing was modality-dependent, with a distinctly steeper gradient for controller direct touch. Third, higher digital literacy was associated with faster TMT-VR completion but not with Fitts task times. Fourth, the psychomotor and cognitive-flexibility measures shared a processing-speed component that accounted for part of the age differences, and the exploratory extension confined this shared component to completion-time and error-adjusted-time indices and the digital-literacy association to error-adjusted time.

\subsection{Interaction modality}\label{sec:4.1}

Task difficulty lawfully increased movement time in both sessions, consistent with Fitts' relation in immersive VR \parencite{ref6,ref7,ref56}. In the direct session, controller touch outperformed the bare virtual finger with a medium-to-large effect (d = 0.67). Because selection was confirmed by steady aim in every mode, this difference cannot reflect different confirmation mechanics; a plausible account is that the physical mass and grip of the controller keep the pointer steady during the dwell period, whereas the bare virtual finger must be held steady in mid-air without a physical anchor and without haptic feedback \parencite{ref8}. In the distant session, controller ray-casting was fastest and eye-gaze slowest, consistent with earlier comparisons of eye-based and controller-based selection in VR \parencite{ref57}. This ordering is the reverse of previous TMT-VR findings, in which eye- and head-gaze outperformed controller-based pointing \parencite{ref26,ref27}, and the reversal is best explained by the different demands of the two tasks.

The TMT-VR is a cognitive task in which searching for the next target is central, so attention, accuracy, and speed all contribute to performance and coupling selection to the visual focus is advantageous \parencite{ref27}; in the Fitts task, attention is guided to a single highlighted target, so performance depends chiefly on psychomotor speed, and the stability of a hand-held pointer prevails. Other HMD comparisons, however, have found eye-gaze pointing to be faster than head pointing and comparable to controller pointing \parencite{ref9,ref10}, and dwell confirmation, although the most accurate hands-free mechanism, is slower than a button press \parencite{ref45}; the standing of eye-gaze relative to head-gaze may therefore depend on the eye-tracking configuration and the confirmation mechanism used. Overall, mode rankings are task-dependent and should not be transferred from one task context to another; the technique should be matched to the demands of the task, and gaze-based techniques remain a viable alternative where prolonged use or motor limitations make upper-limb load a concern.

\subsection{Ageing and the Age \ensuremath{\times} Mode interaction}\label{sec:4.2}

Age was associated with slower task times in every mode, corroborating the motor-ageing literature \parencite{ref17,ref58}. The novel finding is the modality dependence of this slowing across the full 19--90 range: in the common model, controller direct touch carried a distinctly steeper age gradient than every other technique, whereas the remaining four techniques---including both gaze modes and the surface-free virtual finger---did not differ reliably from one another. On the log-seconds scale, estimated slowing was 6.9\% per decade for direct touch against 2.8--5.0\% for the other modes, whereas on raw seconds the interaction was not clear; the modality dependence is therefore a property of relative rather than absolute slowing. Within the distant session, the difficulty-cell analyses showed the same pattern, with ray-casting carrying the steepest age slope (\S{}3.2). The accuracy index showed a complementary pattern: with increasing age, index values decreased for head-gaze and ray-casting but increased for controller direct touch (\S{}3.3). These slopes are small in absolute terms---the index is governed mainly by target width, and within a difficulty cell it varied between participants by only 1--2\% of its value for head-gaze and ray-casting (\S{}3.3)---and because the index combines variability in distance from the target centre and centring, they are read as age-related shifts in the index itself rather than as evidence that distant selection becomes harder with age.

The most age-sensitive technique was therefore mid-air controller touch. This ordering diverges instructively from the non-immersive touch literature, in which direct input attenuated age differences relative to the mouse \parencite{ref12}. Several candidate mechanisms are compatible with the data, and the present design cannot separate them. One is the absence of a physical support surface, which on touch panels anchors and terminates the movement; without it, deceleration and stabilisation of the final pointing position, submovement components affected in older adults \parencite{ref18,ref19}, must be controlled without terminal contact information. This account is incomplete on its own, however, because the virtual finger is equally surface-free yet showed one of the shallower gradients. A second candidate is the precision of depth estimation in stereoscopic near space, which direct reaching loads more heavily than ray-casting \parencite{ref14}. A third is visuomotor calibration to an unfamiliar effector mapping (a held controller acting as a touching implement), which may demand more recalibration from older users than the familiar metaphor of pointing with one's own finger. Distinguishing these accounts requires designs that manipulate surface availability, depth cues, and effector mapping directly.

At the same time, mid-air direct selection is the most naturalistic of the techniques examined: it resembles everyday reaching and touching, which increases task verisimilitude; whether this naturalistic interaction also mitigates the influence of digital literacy and technological competence on performance remains a tentative hypothesis \parencite{ref44}; consistent with it, digital literacy was unrelated to Fitts task times in any mode (\S{}3.3). Overall, the present data establish that controller direct touch carried the steepest relative age gradient, whereas the two gaze techniques, ray-casting, and the virtual finger did not differ reliably from one another; in absolute predicted seconds, however, direct touch remained among the fastest techniques at every representative age (\S{}3.3). For age-inclusive XR design, mid-air direct selection should therefore be evaluated for age sensitivity rather than excluded, and no single technique can be recommended for all users.

\subsection{Digital literacy}\label{sec:4.3}

Digital literacy was not associated with Fitts task times in either session and did not moderate the age association, whereas higher digital literacy was associated with faster TMT-VR completion, an association that was modest in size (\ensuremath{\beta} = \ensuremath{-}0.11; \S{}3.4). In the full-battery extension, adjusted for age and education, only the error-adjusted-time associations survived Holm correction, with the completion-time associations just short of the criterion; digital literacy was unrelated to wrong-target errors, selection distance, or the relative Part B indices, so it reflects temporal efficiency rather than a general improvement of every TMT-VR outcome. This pattern is consistent with the different demands of the two tasks (\S{}4.1): digital familiarity may support task comprehension and strategy on the cognitively demanding, search-based TMT-VR, whereas the naturalistic pointing of the Fitts task may depend little on prior technology experience \parencite{ref44}. Because the two coefficients were not contrasted directly, this task-specific pattern should be treated as a hypothesis for future work.

Nevertheless, the finding aligns with concerns that technology-mediated cognitive assessments partly reflect digital experience \parencite{ref37} and with a technological review and quantitative synthesis indicating that technological competence conditions performance and tolerability in head-mounted-display research \parencite{ref39}. It is also consistent with the computerised-testing literature, in which self-reported computer familiarity relates to speeded test scores \parencite{ref40,ref41} but no more strongly than to their paper-and-pencil equivalents \parencite{ref42}. An alternative reading, supported by a recent meta-analysis that associated everyday technology use with a lower risk of cognitive impairment in later life \parencite{ref43}, is that digital literacy partly indexes cognitive engagement rather than test familiarity; the present design cannot separate these accounts. Two points follow. First, digital literacy should be measured and covaried in XR cognitive assessment, especially among older participants, in whom it varies widely. Second, the DSQ indexes self-reported usage and experience rather than the full construct of technological competence, so the two should not be equated.

\subsection{A shared processing-speed component}\label{sec:4.4}

Four results support the processing-speed account \parencite{ref29} and align with convergent evidence from Fitts and Hick--Hyman tasks \parencite{ref30}, computer-administered trail making \parencite{ref31}, brain-structure--TMT-B associations after speed adjustment \parencite{ref32}, and reaction-time slowing across adulthood \parencite{ref33,ref34,ref59}. First, the TMT-VR Age \ensuremath{\times} Part estimate was small (\ensuremath{\beta} = 0.01 [\ensuremath{-}0.05, 0.06]), consistent with part-general slowing. Second, the Fitts-derived movement-time score predicted TMT-VR performance beyond age and attenuated the age coefficient by 15.7\% in the difficulty-cell analysis and 15.9\% in the composite analysis. Third, TMT-A statistically accounted for 29.4\% of the age association with TMT-B. Fourth, the full-battery extension located the association in completion time and error-adjusted time, not in wrong-target errors, mean selection distance, or the B \ensuremath{-} A, B/A, and normalised difference indices, and no predictor \ensuremath{\times} Part interaction survived correction.

Age nevertheless remained significant after speed adjustment, so residual age-related variance may reflect executive, sensory, or strategic factors. Two boundaries apply. First, the design is cross-sectional and both tasks were completed in the same VR session, so the shared variance is correlational and partly reflects shared method variance; these features prevent causal interpretation. Second, the data do not establish a latent speed factor (\S{}3.5) or an association with every TMT-VR outcome (\S{}3.6). Overall, general psychomotor speed is an important correlate of TMT-VR temporal performance, and a brief psychomotor measure adds information beyond chronological age.

\subsection{Theoretical and design implications}\label{sec:4.5}

The theoretical contribution of this study is the separation of three quantities that are often conflated in XR evaluation: absolute efficiency, relative age sensitivity, and cross-task processing-speed variance. Controller direct touch was fast in absolute terms yet carried the steepest proportional age gradient, whereas eye-gaze showed a shallower age gradient but remained slow in absolute seconds. Age-inclusive design decisions therefore cannot be inferred from young-adult throughput or from an Age \ensuremath{\times} Mode interaction alone; both representative-age predictions and formal slope comparisons are needed, and they remain tied to the hardware, target geometry, and confirmation rule that generated them.

For XR-based assessment, the practical contribution is a two-layer interpretation of TMT-VR scores: the cognitively meaningful outcomes are retained, but they are interpreted alongside a brief motor-speed measure, digital literacy, and the interaction technique used, because all three were associated with performance here. Scores obtained with different techniques should not be converted into one another. Immersive VR assessment can attain convergent validity with established batteries \parencite{ref60} and is sensitive to cognitive impairment \parencite{ref61}, but systematic reviews indicate that few instruments yet report the psychometric properties and normative data required for clinical use \parencite{ref62,ref63}; the present recommendations therefore align with practice parameters for technology-enhanced neuropsychological assessment \parencite{ref64}. Version-controlled administration, mode-specific reference data, and direct testing with older and motor-limited users are the appropriate path from laboratory comparison to inclusive deployment \parencite{ref4,ref15}.

\subsection{Limitations and future directions}\label{sec:4.6}

The within-person five-mode comparison, the common-model slope contrasts, the representative-age predictions, and the full-sample sensitivity analysis are strengths of this study. Its limitations suggest three directions for future research. First, the design was cross-sectional and cannot separate ageing from generational differences in digital experience; longitudinal designs are needed. Second, only three participants were older than 80, so estimates for the oldest decade remain imprecise, and future studies should recruit larger samples of adults over 80. Finally, several analyses were post hoc and exploratory. Replication across devices, confirmation rules, target geometries, and clinical populations is thus needed to establish generalisability.

\section{Conclusions}\label{sec:5}

Interaction modality, age, and their interaction jointly shaped psychomotor performance in immersive VR. Controller direct touch carried a distinctly steeper age gradient than the four other techniques, which did not differ reliably from one another, yet it remained among the fastest techniques in absolute terms at every representative age; the finding therefore motivates age-sensitivity evaluation of mid-air direct selection. Higher digital literacy was associated with faster TMT-VR completion but not with Fitts task times, so digital literacy should be measured and covaried in XR cognitive assessment. The psychomotor and cognitive-flexibility measures shared a processing-speed component that accounted for part of the age differences, and this component was confined to completion-time and error-adjusted-time indices, the digital-literacy association to error-adjusted time; a brief psychomotor measure therefore adds information beyond chronological age when TMT-VR scores are interpreted. Norms and longitudinal comparisons for VR-based assessment should hold interaction mode, target geometry, and confirmation rule constant.

\section*{Supplementary Materials}

The accompanying supplementary document contains Table S1: Exploratory full-battery extension---all 208 likelihood-ratio tests of the Fitts-derived predictors and digital literacy across the 26-variable TMT-VR battery.

\section*{Author Contributions}

Conceptualization, P.K. and K.D.; methodology, P.K. and K.D.; software, P.K.; validation, P.K., K.D., L.A., P.R. and M.R.; formal analysis, P.K.; investigation, P.K., K.D. and L.A.; resources, P.K.; data curation, P.K., K.D. and L.A.; writing---original draft preparation, P.K. and K.D.; writing---review and editing, P.K., K.D., L.A., P.R. and M.R.; visualization, P.K.; supervision, P.K., P.R. and M.R.; project administration, P.K., P.R. and M.R.; funding acquisition, P.K. All authors have read and agreed to the published version of the manuscript.

\section*{Funding}

This project has received funding from the European Union's Horizon Europe research and innovation programme under the Marie Skłodowska-Curie grant agreement No 101149390 (GERIATRIC). Funded by the European Union.

Views and opinions expressed are however those of the author(s) only and do not necessarily reflect those of the European Union or the European Research Executive Agency (REA). Neither the European Union nor the granting authority can be held responsible for them.

\section*{Institutional Review Board Statement}

The study was conducted in accordance with the Declaration of Helsinki. It was carried out within the GERIATRIC programme (Horizon Europe MSCA grant agreement No 101149390) and approved by the Research Ethics and Deontology Committee (\ensuremath{\mathrm{E}}.\ensuremath{\mathrm{H}}.\ensuremath{\Delta}.\ensuremath{\mathrm{E}}.) of the National and Kapodistrian University of Athens (protocol No 20579/5.3.2024; approval decision No 152/4.4.2024) and by the Institutional Review Board of The American College of Greece (expedited review, protocol \#202602564; approved 16 February 2026).

\section*{Informed Consent Statement}

Written informed consent was obtained from all participants involved in the study.

\section*{Data Availability Statement}

The de-identified processed analysis datasets---the person-level file \path{P12_202_public_v03.csv} (202 participants) and the difficulty-cell file \path{FittsTMT_Long_202_public_v02.csv} (4,040 rows), each with its variable codebook---and the analysis code---\path{FINAL_P1_P2_Models_v11_public.R}, which contains the difficulty-cell, person-level, TMT-VR and cross-domain chains and the exploratory full-battery extension (whose machine-readable output is Supplementary Table S1), the descriptive scripts \path{P1_RT_CSQ_Descriptives_c130_v02_public.R} and \path{P1_CSQ_Symptoms_c130_v02_public.R} (reaction time and cybersickness, \S{}3), the accuracy-index decomposition \path{HEISENBERG_AI_CHECK_c130_v02_public.R} (\S{}3.3), and the figure script \path{FIG_P1_Fig3_Fig4_c129_v01_public.R} (Figures 4 and 5), with their complete output logs within the one-command chain \path{RUN_ALL_P12_v13_public.R}---are deposited in the public Zenodo repository of the GERIATRIC programme, limited to what participant consent and the committee approvals permit (\url{https://zenodo.org/records/22767741}; doi: \href{https://doi.org/10.5281/zenodo.22767741}{10.5281/zenodo.22767741}). The deposited datasets are the public versions of the analysis files: the participant identifier is arbitrary and cannot be resolved outside the programme, and the files carry the variables the reported analyses use. Every script verifies the SHA-256 checksum of its input before running, so the results reported here reproduce exactly from the deposited files; the deposited scripts differ from those used in the analysis only in that checksum gate and in their comments, and every difference is enumerated in the repository. The data-assembly scripts that build the datasets from the application exports, and the raw materials, remain stored by the laboratories and are available from the corresponding author.

\section*{Acknowledgments}

The authors thank all participants for their time and engagement. They are grateful to the `Nestor' Psychogeriatric Association and the Amaroussion Open Care Centre for Older Adults (KAPI), through which the older participants were recruited, and to the research assistants of the Psychology Network Lab (PsyNet Lab) of the American College of Greece and of the Experimental Psychology Lab of the National and Kapodistrian University of Athens for their assistance with recruitment, participant scheduling, and data collection.

During the preparation of this work the authors used Anthropic Claude (Claude Fable 5, via Claude Code/Cowork; August--September 2026) and OpenAI Codex (GPT-5.6 Sol Ultra and GPT-6 Astra; August--September 2026) to assist with coding of the R analysis scripts, language refinement, and manuscript editing. The authors reviewed, verified, and edited all outputs and take full responsibility for the content of this work. Generative AI tools are not authors and received no CRediT roles. The built-in image-editing tool in OpenAI Codex was also used to enhance and compose the task illustration in Figure 1 from supplied screenshots and a published figure; the authors reviewed the resulting illustration. The photographs in Figure 3 were edited with the same tool to protect the participants' identity: the background was replaced and faces, body features and personal belongings were blurred; the headset and controller were not edited.

\section*{Conflicts of Interest}

The authors declare no conflicts of interest.

\section*{Author identifiers}
ORCID iDs: P.K., \url{https://orcid.org/0000-0002-2914-1064}; K.D., \url{https://orcid.org/0000-0002-8704-2697}; L.A., \url{https://orcid.org/0009-0004-6770-8852}; P.R., \url{https://orcid.org/0000-0003-1465-2117}; M.R., \url{https://orcid.org/0000-0002-2826-162X}.

\printbibliography[title={References}]
\end{document}


{\noindent\Large\bfseries Supplementary Materials\par}
\vspace{3pt}
{\noindent\large\bfseries Ageing, Digital Literacy, and Interaction Modality in Immersive Virtual Reality: Psychomotor Performance, Cognitive Flexibility, and Their Processing-Speed Association\par}
\vspace{7pt}
\noindent Table S1. Exploratory full-battery extension: all 208 likelihood-ratio tests of the Fitts-derived predictors (182 tests: the difficulty-cell Fitts movement-time score, the five-mode composite, and the five single-mode task times) and digital literacy (26 tests) across the 26-variable TMT-VR battery, all in the same 202 participants.
\setlength{\tabcolsep}{2pt}
\renewcommand{\arraystretch}{1.16}
\fontsize{8}{9.4}\selectfont
\begin{longtable}{L{0.157778\dimexpr\textwidth-52pt\relax}L{0.056667\dimexpr\textwidth-52pt\relax}L{0.146667\dimexpr\textwidth-52pt\relax}L{0.044444\dimexpr\textwidth-52pt\relax}L{0.051111\dimexpr\textwidth-52pt\relax}L{0.100000\dimexpr\textwidth-52pt\relax}L{0.047778\dimexpr\textwidth-52pt\relax}L{0.026667\dimexpr\textwidth-52pt\relax}L{0.051111\dimexpr\textwidth-52pt\relax}L{0.051111\dimexpr\textwidth-52pt\relax}L{0.051111\dimexpr\textwidth-52pt\relax}L{0.166667\dimexpr\textwidth-52pt\relax}L{0.048889\dimexpr\textwidth-52pt\relax}}
\toprule
\textbf{Predictor} & \textbf{Sample} & \textbf{Outcome} & \textbf{Level} & \textbf{Test} & \textbf{Family} & \textbf{\ensuremath{\chi}\textsuperscript{2}} & \textbf{\textit{df}} & \textbf{Raw \textit{p}} & \textbf{Holm \textit{p}} & \textbf{BH \textit{p}} & \textbf{\textit{b} [95\% CI]} & \textbf{Holm sig.} \\
\midrule\endfirsthead
\multicolumn{13}{l}{\textbf{Table S1.} Continued}\\[3pt]\toprule
\textbf{Predictor} & \textbf{Sample} & \textbf{Outcome} & \textbf{Level} & \textbf{Test} & \textbf{Family} & \textbf{\ensuremath{\chi}\textsuperscript{2}} & \textbf{\textit{df}} & \textbf{Raw \textit{p}} & \textbf{Holm \textit{p}} & \textbf{BH \textit{p}} & \textbf{\textit{b} [95\% CI]} & \textbf{Holm sig.} \\
\midrule\endhead
\midrule\multicolumn{13}{r}{\footnotesize Continued on next page}\\\endfoot
\bottomrule\endlastfoot
Five-mode Fitts composite & N = 202 & A/B task time & part & main & Time & 10.32 & 1 & .001 & .005 & .004 & 0.17 [0.07, 0.27] & yes \\
Five-mode Fitts composite & N = 202 & A/B task time & part & \ensuremath{\times} Part & Time & 0.36 & 1 & .550 & .550 & .550 & 0.02 [\ensuremath{-}0.04, 0.08] & no \\
Fitts task time: Head & N = 202 & A/B task time & part & main & Time & 10.29 & 1 & .001 & .025 & .009 & 0.15 [0.06, 0.23] & yes \\
Fitts task time: Head & N = 202 & A/B task time & part & \ensuremath{\times} Part & Time & 0.22 & 1 & .641 & 1.000 & .913 & 0.01 [\ensuremath{-}0.05, 0.07] & no \\
Fitts task time: Eye & N = 202 & A/B task time & part & main & Time & 0.93 & 1 & .335 & 1.000 & .569 & 0.04 [\ensuremath{-}0.04, 0.12] & no \\
Fitts task time: Eye & N = 202 & A/B task time & part & \ensuremath{\times} Part & Time & 0.10 & 1 & .758 & 1.000 & .913 & \ensuremath{-}0.01 [\ensuremath{-}0.07, 0.05] & no \\
Fitts task time: Finger & N = 202 & A/B task time & part & main & Time & 2.79 & 1 & .095 & 1.000 & .247 & 0.07 [\ensuremath{-}0.01, 0.16] & no \\
Fitts task time: Finger & N = 202 & A/B task time & part & \ensuremath{\times} Part & Time & 0.52 & 1 & .472 & 1.000 & .913 & 0.02 [\ensuremath{-}0.04, 0.08] & no \\
Fitts task time: Controller raycast & N = 202 & A/B task time & part & main & Time & 4.14 & 1 & .042 & .671 & .168 & 0.10 [0.00, 0.19] & no \\
Fitts task time: Controller raycast & N = 202 & A/B task time & part & \ensuremath{\times} Part & Time & 0.01 & 1 & .913 & 1.000 & .913 & 0.00 [\ensuremath{-}0.06, 0.06] & no \\
Fitts task time: Controller direct & N = 202 & A/B task time & part & main & Time & 10.78 & 1 & .001 & .021 & .009 & 0.20 [0.08, 0.32] & yes \\
Fitts task time: Controller direct & N = 202 & A/B task time & part & \ensuremath{\times} Part & Time & 1.54 & 1 & .214 & 1.000 & .913 & 0.04 [\ensuremath{-}0.02, 0.10] & no \\
Difficulty-cell Fitts score & N = 202 & A/B task time & part & main & Time & 13.80 & 1 & < .001 & < .001 & < .001 & 0.18 [0.09, 0.28] & yes \\
Difficulty-cell Fitts score & N = 202 & A/B task time & part & \ensuremath{\times} Part & Time & 0.60 & 1 & .440 & .440 & .440 & 0.02 [\ensuremath{-}0.04, 0.08] & no \\
Five-mode Fitts composite & N = 202 & A/B errors & part & main & Errors & 0.59 & 1 & .444 & .933 & .665 & \ensuremath{-}0.08 [\ensuremath{-}0.27, 0.12] & no \\
Five-mode Fitts composite & N = 202 & A/B errors & part & \ensuremath{\times} Part & Errors & 1.34 & 1 & .247 & .247 & .247 & 0.07 [\ensuremath{-}0.05, 0.18] & no \\
Fitts task time: Head & N = 202 & A/B errors & part & main & Errors & 0.18 & 1 & .674 & 1.000 & .879 & \ensuremath{-}0.04 [\ensuremath{-}0.20, 0.13] & no \\
Fitts task time: Head & N = 202 & A/B errors & part & \ensuremath{\times} Part & Errors & 0.62 & 1 & .433 & 1.000 & .709 & 0.04 [\ensuremath{-}0.07, 0.16] & no \\
Fitts task time: Eye & N = 202 & A/B errors & part & main & Errors & 1.07 & 1 & .301 & 1.000 & .879 & \ensuremath{-}0.08 [\ensuremath{-}0.24, 0.07] & no \\
Fitts task time: Eye & N = 202 & A/B errors & part & \ensuremath{\times} Part & Errors & 0.33 & 1 & .567 & 1.000 & .709 & \ensuremath{-}0.03 [\ensuremath{-}0.15, 0.08] & no \\
Fitts task time: Finger & N = 202 & A/B errors & part & main & Errors & 0.23 & 1 & .631 & 1.000 & .879 & \ensuremath{-}0.04 [\ensuremath{-}0.20, 0.12] & no \\
Fitts task time: Finger & N = 202 & A/B errors & part & \ensuremath{\times} Part & Errors & 6.07 & 1 & .014 & .069 & .069 & 0.15 [0.03, 0.26] & no \\
Fitts task time: Controller raycast & N = 202 & A/B errors & part & main & Errors & 0.24 & 1 & .625 & 1.000 & .879 & \ensuremath{-}0.04 [\ensuremath{-}0.22, 0.13] & no \\
Fitts task time: Controller raycast & N = 202 & A/B errors & part & \ensuremath{\times} Part & Errors & 0.06 & 1 & .808 & 1.000 & .808 & 0.01 [\ensuremath{-}0.10, 0.12] & no \\
Fitts task time: Controller direct & N = 202 & A/B errors & part & main & Errors & 0.01 & 1 & .934 & 1.000 & .934 & 0.01 [\ensuremath{-}0.22, 0.24] & no \\
Fitts task time: Controller direct & N = 202 & A/B errors & part & \ensuremath{\times} Part & Errors & 2.30 & 1 & .130 & .519 & .324 & 0.08 [\ensuremath{-}0.03, 0.19] & no \\
Difficulty-cell Fitts score & N = 202 & A/B errors & part & main & Errors & 1.30 & 1 & .255 & .509 & .382 & \ensuremath{-}0.11 [\ensuremath{-}0.29, 0.08] & no \\
Difficulty-cell Fitts score & N = 202 & A/B errors & part & \ensuremath{\times} Part & Errors & 0.41 & 1 & .524 & .524 & .524 & 0.04 [\ensuremath{-}0.07, 0.14] & no \\
Five-mode Fitts composite & N = 202 & A/B selection distance & part & main & Selection distance & 0.49 & 1 & .483 & 1.000 & .591 & 0.05 [\ensuremath{-}0.10, 0.21] & no \\
Five-mode Fitts composite & N = 202 & A/B selection distance & part & \ensuremath{\times} Part & Selection distance & 0.58 & 1 & .445 & .445 & .445 & 0.03 [\ensuremath{-}0.04, 0.10] & no \\
Fitts task time: Head & N = 202 & A/B selection distance & part & main & Selection distance & 1.64 & 1 & .200 & 1.000 & .457 & 0.09 [\ensuremath{-}0.05, 0.22] & no \\
Fitts task time: Head & N = 202 & A/B selection distance & part & \ensuremath{\times} Part & Selection distance & 0.58 & 1 & .447 & .934 & .559 & 0.03 [\ensuremath{-}0.04, 0.10] & no \\
Fitts task time: Eye & N = 202 & A/B selection distance & part & main & Selection distance & 0.13 & 1 & .722 & 1.000 & .773 & \ensuremath{-}0.02 [\ensuremath{-}0.14, 0.10] & no \\
Fitts task time: Eye & N = 202 & A/B selection distance & part & \ensuremath{\times} Part & Selection distance & 1.57 & 1 & .210 & .842 & .519 & \ensuremath{-}0.04 [\ensuremath{-}0.11, 0.03] & no \\
Fitts task time: Finger & N = 202 & A/B selection distance & part & main & Selection distance & 0.49 & 1 & .483 & 1.000 & .659 & \ensuremath{-}0.04 [\ensuremath{-}0.17, 0.08] & no \\
Fitts task time: Finger & N = 202 & A/B selection distance & part & \ensuremath{\times} Part & Selection distance & 3.56 & 1 & .059 & .296 & .296 & 0.07 [\ensuremath{-}0.00, 0.14] & no \\
Fitts task time: Controller raycast & N = 202 & A/B selection distance & part & main & Selection distance & 1.84 & 1 & .175 & 1.000 & .457 & 0.10 [\ensuremath{-}0.04, 0.24] & no \\
Fitts task time: Controller raycast & N = 202 & A/B selection distance & part & \ensuremath{\times} Part & Selection distance & 1.03 & 1 & .311 & .934 & .519 & 0.04 [\ensuremath{-}0.03, 0.11] & no \\
Fitts task time: Controller direct & N = 202 & A/B selection distance & part & main & Selection distance & 1.35 & 1 & .245 & 1.000 & .457 & 0.11 [\ensuremath{-}0.07, 0.28] & no \\
Fitts task time: Controller direct & N = 202 & A/B selection distance & part & \ensuremath{\times} Part & Selection distance & 0.19 & 1 & .662 & .934 & .662 & 0.02 [\ensuremath{-}0.05, 0.09] & no \\
Difficulty-cell Fitts score & N = 202 & A/B selection distance & part & main & Selection distance & 0.03 & 1 & .860 & 1.000 & .911 & \ensuremath{-}0.01 [\ensuremath{-}0.16, 0.13] & no \\
Difficulty-cell Fitts score & N = 202 & A/B selection distance & part & \ensuremath{\times} Part & Selection distance & 0.05 & 1 & .817 & .817 & .817 & 0.01 [\ensuremath{-}0.06, 0.08] & no \\
Five-mode Fitts composite & N = 202 & A/B adjusted time & part & main & Adjusted time & 8.39 & 1 & .004 & .011 & .011 & 0.15 [0.05, 0.25] & yes \\
Five-mode Fitts composite & N = 202 & A/B adjusted time & part & \ensuremath{\times} Part & Adjusted time & 0.84 & 1 & .361 & .361 & .361 & 0.03 [\ensuremath{-}0.03, 0.09] & no \\
Fitts task time: Head & N = 202 & A/B adjusted time & part & main & Adjusted time & 8.97 & 1 & .003 & .038 & .020 & 0.14 [0.05, 0.22] & yes \\
Fitts task time: Head & N = 202 & A/B adjusted time & part & \ensuremath{\times} Part & Adjusted time & 0.40 & 1 & .528 & 1.000 & .785 & 0.02 [\ensuremath{-}0.04, 0.08] & no \\
Fitts task time: Eye & N = 202 & A/B adjusted time & part & main & Adjusted time & 0.43 & 1 & .512 & 1.000 & .639 & 0.03 [\ensuremath{-}0.05, 0.11] & no \\
Fitts task time: Eye & N = 202 & A/B adjusted time & part & \ensuremath{\times} Part & Adjusted time & 0.08 & 1 & .775 & 1.000 & .785 & \ensuremath{-}0.01 [\ensuremath{-}0.07, 0.05] & no \\
Fitts task time: Finger & N = 202 & A/B adjusted time & part & main & Adjusted time & 2.48 & 1 & .115 & 1.000 & .288 & 0.07 [\ensuremath{-}0.02, 0.15] & no \\
Fitts task time: Finger & N = 202 & A/B adjusted time & part & \ensuremath{\times} Part & Adjusted time & 1.68 & 1 & .195 & .781 & .488 & 0.04 [\ensuremath{-}0.02, 0.10] & no \\
Fitts task time: Controller raycast & N = 202 & A/B adjusted time & part & main & Adjusted time & 3.37 & 1 & .066 & .729 & .199 & 0.09 [\ensuremath{-}0.01, 0.18] & no \\
Fitts task time: Controller raycast & N = 202 & A/B adjusted time & part & \ensuremath{\times} Part & Adjusted time & 0.07 & 1 & .785 & 1.000 & .785 & 0.01 [\ensuremath{-}0.05, 0.07] & no \\
Fitts task time: Controller direct & N = 202 & A/B adjusted time & part & main & Adjusted time & 9.32 & 1 & .002 & .034 & .020 & 0.19 [0.07, 0.31] & yes \\
Fitts task time: Controller direct & N = 202 & A/B adjusted time & part & \ensuremath{\times} Part & Adjusted time & 2.22 & 1 & .137 & .683 & .488 & 0.04 [\ensuremath{-}0.01, 0.10] & no \\
Difficulty-cell Fitts score & N = 202 & A/B adjusted time & part & main & Adjusted time & 10.28 & 1 & .001 & .004 & .003 & 0.16 [0.06, 0.25] & yes \\
Difficulty-cell Fitts score & N = 202 & A/B adjusted time & part & \ensuremath{\times} Part & Adjusted time & 1.02 & 1 & .313 & .313 & .313 & 0.03 [\ensuremath{-}0.03, 0.09] & no \\
Five-mode Fitts composite & N = 202 & A/B error rate & part & main & Algebraic re-expression & 0.29 & 1 & .591 & 1.000 & .793 & \ensuremath{-}0.04 [\ensuremath{-}0.17, 0.09] & no \\
Five-mode Fitts composite & N = 202 & A/B error rate & part & \ensuremath{\times} Part & Algebraic re-expression & 0.34 & 1 & .560 & 1.000 & .669 & 0.02 [\ensuremath{-}0.05, 0.10] & no \\
Fitts task time: Head & N = 202 & A/B error rate & part & main & Algebraic re-expression & 0.14 & 1 & .707 & 1.000 & .993 & \ensuremath{-}0.02 [\ensuremath{-}0.14, 0.09] & no \\
Fitts task time: Head & N = 202 & A/B error rate & part & \ensuremath{\times} Part & Algebraic re-expression & 0.07 & 1 & .789 & 1.000 & .904 & 0.01 [\ensuremath{-}0.07, 0.09] & no \\
Fitts task time: Eye & N = 202 & A/B error rate & part & main & Algebraic re-expression & 0.81 & 1 & .369 & 1.000 & .993 & \ensuremath{-}0.05 [\ensuremath{-}0.15, 0.06] & no \\
Fitts task time: Eye & N = 202 & A/B error rate & part & \ensuremath{\times} Part & Algebraic re-expression & 0.33 & 1 & .566 & 1.000 & .904 & \ensuremath{-}0.02 [\ensuremath{-}0.10, 0.05] & no \\
Fitts task time: Finger & N = 202 & A/B error rate & part & main & Algebraic re-expression & 0.19 & 1 & .663 & 1.000 & .993 & \ensuremath{-}0.02 [\ensuremath{-}0.13, 0.08] & no \\
Fitts task time: Finger & N = 202 & A/B error rate & part & \ensuremath{\times} Part & Algebraic re-expression & 4.83 & 1 & .028 & .279 & .267 & 0.08 [0.01, 0.16] & no \\
Fitts task time: Controller raycast & N = 202 & A/B error rate & part & main & Algebraic re-expression & 0.05 & 1 & .821 & 1.000 & .993 & \ensuremath{-}0.01 [\ensuremath{-}0.13, 0.11] & no \\
Fitts task time: Controller raycast & N = 202 & A/B error rate & part & \ensuremath{\times} Part & Algebraic re-expression & 0.03 & 1 & .868 & 1.000 & .904 & \ensuremath{-}0.01 [\ensuremath{-}0.08, 0.07] & no \\
Fitts task time: Controller direct & N = 202 & A/B error rate & part & main & Algebraic re-expression & 0.15 & 1 & .696 & 1.000 & .993 & 0.03 [\ensuremath{-}0.12, 0.18] & no \\
Fitts task time: Controller direct & N = 202 & A/B error rate & part & \ensuremath{\times} Part & Algebraic re-expression & 0.21 & 1 & .649 & 1.000 & .904 & 0.02 [\ensuremath{-}0.06, 0.09] & no \\
Difficulty-cell Fitts score & N = 202 & A/B error rate & part & main & Algebraic re-expression & 1.02 & 1 & .313 & 1.000 & .874 & \ensuremath{-}0.06 [\ensuremath{-}0.19, 0.06] & no \\
Difficulty-cell Fitts score & N = 202 & A/B error rate & part & \ensuremath{\times} Part & Algebraic re-expression & 0.03 & 1 & .854 & 1.000 & .961 & 0.01 [\ensuremath{-}0.07, 0.08] & no \\
Five-mode Fitts composite & N = 202 & A/B accuracy & part & main & Algebraic re-expression & 0.25 & 1 & .616 & 1.000 & .793 & 0.03 [\ensuremath{-}0.10, 0.16] & no \\
Five-mode Fitts composite & N = 202 & A/B accuracy & part & \ensuremath{\times} Part & Algebraic re-expression & 0.18 & 1 & .669 & 1.000 & .669 & \ensuremath{-}0.02 [\ensuremath{-}0.09, 0.06] & no \\
Fitts task time: Head & N = 202 & A/B accuracy & part & main & Algebraic re-expression & 0.10 & 1 & .752 & 1.000 & .993 & 0.02 [\ensuremath{-}0.10, 0.13] & no \\
Fitts task time: Head & N = 202 & A/B accuracy & part & \ensuremath{\times} Part & Algebraic re-expression & 0.05 & 1 & .827 & 1.000 & .904 & \ensuremath{-}0.01 [\ensuremath{-}0.08, 0.07] & no \\
Fitts task time: Eye & N = 202 & A/B accuracy & part & main & Algebraic re-expression & 0.79 & 1 & .373 & 1.000 & .993 & 0.05 [\ensuremath{-}0.06, 0.15] & no \\
Fitts task time: Eye & N = 202 & A/B accuracy & part & \ensuremath{\times} Part & Algebraic re-expression & 0.59 & 1 & .443 & 1.000 & .904 & 0.03 [\ensuremath{-}0.05, 0.11] & no \\
Fitts task time: Finger & N = 202 & A/B accuracy & part & main & Algebraic re-expression & 0.16 & 1 & .689 & 1.000 & .993 & 0.02 [\ensuremath{-}0.09, 0.13] & no \\
Fitts task time: Finger & N = 202 & A/B accuracy & part & \ensuremath{\times} Part & Algebraic re-expression & 3.73 & 1 & .053 & .481 & .267 & \ensuremath{-}0.07 [\ensuremath{-}0.15, 0.00] & no \\
Fitts task time: Controller raycast & N = 202 & A/B accuracy & part & main & Algebraic re-expression & 0.07 & 1 & .791 & 1.000 & .993 & 0.02 [\ensuremath{-}0.10, 0.14] & no \\
Fitts task time: Controller raycast & N = 202 & A/B accuracy & part & \ensuremath{\times} Part & Algebraic re-expression & 0.01 & 1 & .904 & 1.000 & .904 & 0.00 [\ensuremath{-}0.07, 0.08] & no \\
Fitts task time: Controller direct & N = 202 & A/B accuracy & part & main & Algebraic re-expression & 0.21 & 1 & .645 & 1.000 & .993 & \ensuremath{-}0.04 [\ensuremath{-}0.19, 0.12] & no \\
Fitts task time: Controller direct & N = 202 & A/B accuracy & part & \ensuremath{\times} Part & Algebraic re-expression & 0.11 & 1 & .736 & 1.000 & .904 & \ensuremath{-}0.01 [\ensuremath{-}0.09, 0.06] & no \\
Difficulty-cell Fitts score & N = 202 & A/B accuracy & part & main & Algebraic re-expression & 0.92 & 1 & .338 & 1.000 & .874 & 0.06 [\ensuremath{-}0.06, 0.18] & no \\
Difficulty-cell Fitts score & N = 202 & A/B accuracy & part & \ensuremath{\times} Part & Algebraic re-expression & 0.00 & 1 & .961 & 1.000 & .961 & 0.00 [\ensuremath{-}0.07, 0.08] & no \\
Five-mode Fitts composite & N = 202 & total time & person & main & Time & 9.76 & 1 & .002 & .005 & .004 & 0.19 [0.07, 0.31] & yes \\
Fitts task time: Head & N = 202 & total time & person & main & Time & 10.22 & 1 & .001 & .025 & .009 & 0.17 [0.07, 0.28] & yes \\
Fitts task time: Eye & N = 202 & total time & person & main & Time & 1.75 & 1 & .186 & 1.000 & .414 & 0.07 [\ensuremath{-}0.03, 0.16] & no \\
Fitts task time: Finger & N = 202 & total time & person & main & Time & 2.73 & 1 & .099 & 1.000 & .247 & 0.08 [\ensuremath{-}0.02, 0.18] & no \\
Fitts task time: Controller raycast & N = 202 & total time & person & main & Time & 2.92 & 1 & .088 & 1.000 & .247 & 0.10 [\ensuremath{-}0.01, 0.21] & no \\
Fitts task time: Controller direct & N = 202 & total time & person & main & Time & 7.95 & 1 & .005 & .082 & .024 & 0.21 [0.06, 0.35] & no \\
Difficulty-cell Fitts score & N = 202 & total time & person & main & Time & 13.07 & 1 & < .001 & < .001 & < .001 & 0.21 [0.10, 0.32] & yes \\
Five-mode Fitts composite & N = 202 & total errors & person & main & Errors & 1.03 & 1 & .311 & .933 & .665 & \ensuremath{-}0.10 [\ensuremath{-}0.29, 0.09] & no \\
Fitts task time: Head & N = 202 & total errors & person & main & Errors & 0.34 & 1 & .558 & 1.000 & .879 & \ensuremath{-}0.05 [\ensuremath{-}0.21, 0.11] & no \\
Fitts task time: Eye & N = 202 & total errors & person & main & Errors & 0.96 & 1 & .327 & 1.000 & .879 & \ensuremath{-}0.08 [\ensuremath{-}0.23, 0.08] & no \\
Fitts task time: Finger & N = 202 & total errors & person & main & Errors & 0.29 & 1 & .589 & 1.000 & .879 & \ensuremath{-}0.04 [\ensuremath{-}0.20, 0.12] & no \\
Fitts task time: Controller raycast & N = 202 & total errors & person & main & Errors & 0.58 & 1 & .446 & 1.000 & .879 & \ensuremath{-}0.07 [\ensuremath{-}0.24, 0.10] & no \\
Fitts task time: Controller direct & N = 202 & total errors & person & main & Errors & 0.15 & 1 & .703 & 1.000 & .879 & \ensuremath{-}0.04 [\ensuremath{-}0.26, 0.18] & no \\
Difficulty-cell Fitts score & N = 202 & total errors & person & main & Errors & 2.05 & 1 & .152 & .456 & .382 & \ensuremath{-}0.13 [\ensuremath{-}0.30, 0.05] & no \\
Five-mode Fitts composite & N = 202 & total selection distance & person & main & Selection distance & 0.65 & 1 & .421 & 1.000 & .591 & 0.07 [\ensuremath{-}0.10, 0.25] & no \\
Fitts task time: Head & N = 202 & total selection distance & person & main & Selection distance & 1.95 & 1 & .163 & 1.000 & .457 & 0.11 [\ensuremath{-}0.05, 0.26] & no \\
Fitts task time: Eye & N = 202 & total selection distance & person & main & Selection distance & 0.05 & 1 & .828 & 1.000 & .828 & \ensuremath{-}0.02 [\ensuremath{-}0.16, 0.13] & no \\
Fitts task time: Finger & N = 202 & total selection distance & person & main & Selection distance & 0.26 & 1 & .613 & 1.000 & .761 & \ensuremath{-}0.04 [\ensuremath{-}0.18, 0.11] & no \\
Fitts task time: Controller raycast & N = 202 & total selection distance & person & main & Selection distance & 1.42 & 1 & .233 & 1.000 & .457 & 0.10 [\ensuremath{-}0.06, 0.26] & no \\
Fitts task time: Controller direct & N = 202 & total selection distance & person & main & Selection distance & 1.40 & 1 & .237 & 1.000 & .457 & 0.13 [\ensuremath{-}0.08, 0.33] & no \\
Difficulty-cell Fitts score & N = 202 & total selection distance & person & main & Selection distance & 0.01 & 1 & .911 & 1.000 & .911 & \ensuremath{-}0.01 [\ensuremath{-}0.18, 0.16] & no \\
Five-mode Fitts composite & N = 202 & total adjusted time & person & main & Adjusted time & 7.17 & 1 & .007 & .015 & .011 & 0.17 [0.04, 0.29] & yes \\
Fitts task time: Head & N = 202 & total adjusted time & person & main & Adjusted time & 8.32 & 1 & .004 & .051 & .020 & 0.16 [0.05, 0.26] & no \\
Fitts task time: Eye & N = 202 & total adjusted time & person & main & Adjusted time & 1.18 & 1 & .278 & 1.000 & .454 & 0.05 [\ensuremath{-}0.04, 0.15] & no \\
Fitts task time: Finger & N = 202 & total adjusted time & person & main & Adjusted time & 1.93 & 1 & .165 & 1.000 & .321 & 0.07 [\ensuremath{-}0.03, 0.17] & no \\
Fitts task time: Controller raycast & N = 202 & total adjusted time & person & main & Adjusted time & 1.87 & 1 & .171 & 1.000 & .321 & 0.08 [\ensuremath{-}0.03, 0.19] & no \\
Fitts task time: Controller direct & N = 202 & total adjusted time & person & main & Adjusted time & 6.09 & 1 & .014 & .163 & .051 & 0.18 [0.04, 0.33] & no \\
Difficulty-cell Fitts score & N = 202 & total adjusted time & person & main & Adjusted time & 9.37 & 1 & .002 & .004 & .003 & 0.18 [0.07, 0.29] & yes \\
Five-mode Fitts composite & N = 202 & difference time & person & main & Time & 0.61 & 1 & .436 & .872 & .581 & 0.07 [\ensuremath{-}0.10, 0.24] & no \\
Fitts task time: Head & N = 202 & difference time & person & main & Time & 0.24 & 1 & .624 & 1.000 & .780 & 0.04 [\ensuremath{-}0.11, 0.19] & no \\
Fitts task time: Eye & N = 202 & difference time & person & main & Time & 1.03 & 1 & .309 & 1.000 & .569 & 0.07 [\ensuremath{-}0.07, 0.21] & no \\
Fitts task time: Finger & N = 202 & difference time & person & main & Time & 0.01 & 1 & .940 & 1.000 & .940 & \ensuremath{-}0.01 [\ensuremath{-}0.15, 0.14] & no \\
Fitts task time: Controller raycast & N = 202 & difference time & person & main & Time & 0.90 & 1 & .342 & 1.000 & .569 & 0.08 [\ensuremath{-}0.08, 0.24] & no \\
Fitts task time: Controller direct & N = 202 & difference time & person & main & Time & 0.02 & 1 & .881 & 1.000 & .927 & 0.02 [\ensuremath{-}0.19, 0.22] & no \\
Difficulty-cell Fitts score & N = 202 & difference time & person & main & Time & 0.73 & 1 & .394 & .788 & .525 & 0.07 [\ensuremath{-}0.09, 0.23] & no \\
Five-mode Fitts composite & N = 202 & difference errors & person & main & Errors & 0.13 & 1 & .718 & .933 & .718 & \ensuremath{-}0.03 [\ensuremath{-}0.21, 0.15] & no \\
Fitts task time: Head & N = 202 & difference errors & person & main & Errors & 0.05 & 1 & .823 & 1.000 & .901 & \ensuremath{-}0.02 [\ensuremath{-}0.17, 0.14] & no \\
Fitts task time: Eye & N = 202 & difference errors & person & main & Errors & 0.48 & 1 & .488 & 1.000 & .879 & 0.05 [\ensuremath{-}0.09, 0.19] & no \\
Fitts task time: Finger & N = 202 & difference errors & person & main & Errors & 3.70 & 1 & .055 & .818 & .818 & \ensuremath{-}0.14 [\ensuremath{-}0.29, 0.00] & no \\
Fitts task time: Controller raycast & N = 202 & difference errors & person & main & Errors & 0.26 & 1 & .608 & 1.000 & .879 & 0.04 [\ensuremath{-}0.12, 0.21] & no \\
Fitts task time: Controller direct & N = 202 & difference errors & person & main & Errors & 0.04 & 1 & .841 & 1.000 & .901 & \ensuremath{-}0.02 [\ensuremath{-}0.23, 0.19] & no \\
Difficulty-cell Fitts score & N = 202 & difference errors & person & main & Errors & 0.01 & 1 & .943 & .943 & .943 & 0.01 [\ensuremath{-}0.16, 0.18] & no \\
Five-mode Fitts composite & N = 202 & difference selection distance & person & main & Selection distance & 0.29 & 1 & .591 & 1.000 & .591 & \ensuremath{-}0.05 [\ensuremath{-}0.23, 0.13] & no \\
Fitts task time: Head & N = 202 & difference selection distance & person & main & Selection distance & 0.50 & 1 & .478 & 1.000 & .659 & \ensuremath{-}0.06 [\ensuremath{-}0.21, 0.10] & no \\
Fitts task time: Eye & N = 202 & difference selection distance & person & main & Selection distance & 2.11 & 1 & .147 & 1.000 & .457 & 0.11 [\ensuremath{-}0.04, 0.25] & no \\
Fitts task time: Finger & N = 202 & difference selection distance & person & main & Selection distance & 1.20 & 1 & .274 & 1.000 & .457 & \ensuremath{-}0.08 [\ensuremath{-}0.23, 0.07] & no \\
Fitts task time: Controller raycast & N = 202 & difference selection distance & person & main & Selection distance & 1.44 & 1 & .230 & 1.000 & .457 & \ensuremath{-}0.10 [\ensuremath{-}0.27, 0.07] & no \\
Fitts task time: Controller direct & N = 202 & difference selection distance & person & main & Selection distance & 0.19 & 1 & .660 & 1.000 & .761 & \ensuremath{-}0.05 [\ensuremath{-}0.26, 0.16] & no \\
Difficulty-cell Fitts score & N = 202 & difference selection distance & person & main & Selection distance & 0.02 & 1 & .892 & 1.000 & .911 & 0.01 [\ensuremath{-}0.16, 0.18] & no \\
Five-mode Fitts composite & N = 202 & difference adjusted time & person & main & Adjusted time & 0.31 & 1 & .576 & .576 & .576 & 0.05 [\ensuremath{-}0.12, 0.22] & no \\
Fitts task time: Head & N = 202 & difference adjusted time & person & main & Adjusted time & 0.10 & 1 & .755 & 1.000 & .809 & 0.02 [\ensuremath{-}0.13, 0.17] & no \\
Fitts task time: Eye & N = 202 & difference adjusted time & person & main & Adjusted time & 1.06 & 1 & .303 & 1.000 & .454 & 0.07 [\ensuremath{-}0.07, 0.21] & no \\
Fitts task time: Finger & N = 202 & difference adjusted time & person & main & Adjusted time & 0.17 & 1 & .678 & 1.000 & .782 & \ensuremath{-}0.03 [\ensuremath{-}0.17, 0.11] & no \\
Fitts task time: Controller raycast & N = 202 & difference adjusted time & person & main & Adjusted time & 0.72 & 1 & .395 & 1.000 & .538 & 0.07 [\ensuremath{-}0.09, 0.23] & no \\
Fitts task time: Controller direct & N = 202 & difference adjusted time & person & main & Adjusted time & 0.00 & 1 & .980 & 1.000 & .980 & 0.00 [\ensuremath{-}0.20, 0.21] & no \\
Difficulty-cell Fitts score & N = 202 & difference adjusted time & person & main & Adjusted time & 0.45 & 1 & .504 & .504 & .504 & 0.06 [\ensuremath{-}0.11, 0.22] & no \\
Five-mode Fitts composite & N = 202 & ratio time & person & main & Time & 0.09 & 1 & .764 & .872 & .764 & 0.03 [\ensuremath{-}0.15, 0.21] & no \\
Fitts task time: Head & N = 202 & ratio time & person & main & Time & 0.09 & 1 & .761 & 1.000 & .895 & 0.02 [\ensuremath{-}0.13, 0.18] & no \\
Fitts task time: Eye & N = 202 & ratio time & person & main & Time & 0.58 & 1 & .445 & 1.000 & .685 & 0.06 [\ensuremath{-}0.09, 0.20] & no \\
Fitts task time: Finger & N = 202 & ratio time & person & main & Time & 0.02 & 1 & .879 & 1.000 & .927 & \ensuremath{-}0.01 [\ensuremath{-}0.16, 0.14] & no \\
Fitts task time: Controller raycast & N = 202 & ratio time & person & main & Time & 0.28 & 1 & .595 & 1.000 & .780 & 0.04 [\ensuremath{-}0.12, 0.21] & no \\
Fitts task time: Controller direct & N = 202 & ratio time & person & main & Time & 0.46 & 1 & .496 & 1.000 & .709 & \ensuremath{-}0.07 [\ensuremath{-}0.28, 0.14] & no \\
Difficulty-cell Fitts score & N = 202 & ratio time & person & main & Time & 0.08 & 1 & .771 & .788 & .771 & 0.02 [\ensuremath{-}0.14, 0.19] & no \\
Five-mode Fitts composite & N = 202 & total error rate & person & main & Algebraic re-expression & 0.07 & 1 & .793 & 1.000 & .793 & \ensuremath{-}0.02 [\ensuremath{-}0.19, 0.15] & no \\
Fitts task time: Head & N = 202 & total error rate & person & main & Algebraic re-expression & 0.01 & 1 & .917 & 1.000 & .993 & \ensuremath{-}0.01 [\ensuremath{-}0.16, 0.14] & no \\
Fitts task time: Eye & N = 202 & total error rate & person & main & Algebraic re-expression & 0.58 & 1 & .446 & 1.000 & .993 & \ensuremath{-}0.05 [\ensuremath{-}0.19, 0.08] & no \\
Fitts task time: Finger & N = 202 & total error rate & person & main & Algebraic re-expression & 0.19 & 1 & .661 & 1.000 & .993 & \ensuremath{-}0.03 [\ensuremath{-}0.17, 0.11] & no \\
Fitts task time: Controller raycast & N = 202 & total error rate & person & main & Algebraic re-expression & 0.01 & 1 & .936 & 1.000 & .993 & 0.01 [\ensuremath{-}0.15, 0.16] & no \\
Fitts task time: Controller direct & N = 202 & total error rate & person & main & Algebraic re-expression & 0.42 & 1 & .518 & 1.000 & .993 & 0.07 [\ensuremath{-}0.13, 0.26] & no \\
Difficulty-cell Fitts score & N = 202 & total error rate & person & main & Algebraic re-expression & 0.29 & 1 & .591 & 1.000 & .874 & \ensuremath{-}0.04 [\ensuremath{-}0.20, 0.12] & no \\
Five-mode Fitts composite & N = 202 & total accuracy & person & main & Algebraic re-expression & 0.07 & 1 & .789 & 1.000 & .793 & 0.02 [\ensuremath{-}0.14, 0.19] & no \\
Fitts task time: Head & N = 202 & total accuracy & person & main & Algebraic re-expression & 0.00 & 1 & .996 & 1.000 & .996 & \ensuremath{-}0.00 [\ensuremath{-}0.15, 0.15] & no \\
Fitts task time: Eye & N = 202 & total accuracy & person & main & Algebraic re-expression & 0.67 & 1 & .411 & 1.000 & .993 & 0.06 [\ensuremath{-}0.08, 0.19] & no \\
Fitts task time: Finger & N = 202 & total accuracy & person & main & Algebraic re-expression & 0.16 & 1 & .693 & 1.000 & .993 & 0.03 [\ensuremath{-}0.11, 0.16] & no \\
Fitts task time: Controller raycast & N = 202 & total accuracy & person & main & Algebraic re-expression & 0.00 & 1 & .976 & 1.000 & .996 & \ensuremath{-}0.00 [\ensuremath{-}0.16, 0.15] & no \\
Fitts task time: Controller direct & N = 202 & total accuracy & person & main & Algebraic re-expression & 0.32 & 1 & .574 & 1.000 & .993 & \ensuremath{-}0.06 [\ensuremath{-}0.25, 0.14] & no \\
Difficulty-cell Fitts score & N = 202 & total accuracy & person & main & Algebraic re-expression & 0.44 & 1 & .509 & 1.000 & .874 & 0.05 [\ensuremath{-}0.10, 0.21] & no \\
Five-mode Fitts composite & N = 202 & difference error rate & person & main & Algebraic re-expression & 0.09 & 1 & .768 & 1.000 & .793 & \ensuremath{-}0.03 [\ensuremath{-}0.21, 0.15] & no \\
Fitts task time: Head & N = 202 & difference error rate & person & main & Algebraic re-expression & 0.03 & 1 & .861 & 1.000 & .993 & \ensuremath{-}0.01 [\ensuremath{-}0.17, 0.14] & no \\
Fitts task time: Eye & N = 202 & difference error rate & person & main & Algebraic re-expression & 0.54 & 1 & .464 & 1.000 & .993 & 0.05 [\ensuremath{-}0.09, 0.20] & no \\
Fitts task time: Finger & N = 202 & difference error rate & person & main & Algebraic re-expression & 3.60 & 1 & .058 & 1.000 & .993 & \ensuremath{-}0.14 [\ensuremath{-}0.29, 0.00] & no \\
Fitts task time: Controller raycast & N = 202 & difference error rate & person & main & Algebraic re-expression & 0.29 & 1 & .593 & 1.000 & .993 & 0.04 [\ensuremath{-}0.12, 0.21] & no \\
Fitts task time: Controller direct & N = 202 & difference error rate & person & main & Algebraic re-expression & 0.01 & 1 & .920 & 1.000 & .993 & \ensuremath{-}0.01 [\ensuremath{-}0.22, 0.20] & no \\
Difficulty-cell Fitts score & N = 202 & difference error rate & person & main & Algebraic re-expression & 0.03 & 1 & .874 & 1.000 & .874 & 0.01 [\ensuremath{-}0.16, 0.18] & no \\
Five-mode Fitts composite & N = 202 & difference accuracy & person & main & Algebraic re-expression & 0.10 & 1 & .749 & 1.000 & .793 & 0.03 [\ensuremath{-}0.15, 0.21] & no \\
Fitts task time: Head & N = 202 & difference accuracy & person & main & Algebraic re-expression & 0.02 & 1 & .875 & 1.000 & .993 & 0.01 [\ensuremath{-}0.14, 0.17] & no \\
Fitts task time: Eye & N = 202 & difference accuracy & person & main & Algebraic re-expression & 0.47 & 1 & .491 & 1.000 & .993 & \ensuremath{-}0.05 [\ensuremath{-}0.19, 0.09] & no \\
Fitts task time: Finger & N = 202 & difference accuracy & person & main & Algebraic re-expression & 3.75 & 1 & .053 & 1.000 & .993 & 0.14 [\ensuremath{-}0.00, 0.29] & no \\
Fitts task time: Controller raycast & N = 202 & difference accuracy & person & main & Algebraic re-expression & 0.33 & 1 & .566 & 1.000 & .993 & \ensuremath{-}0.05 [\ensuremath{-}0.21, 0.12] & no \\
Fitts task time: Controller direct & N = 202 & difference accuracy & person & main & Algebraic re-expression & 0.03 & 1 & .873 & 1.000 & .993 & 0.02 [\ensuremath{-}0.19, 0.23] & no \\
Difficulty-cell Fitts score & N = 202 & difference accuracy & person & main & Algebraic re-expression & 0.03 & 1 & .872 & 1.000 & .874 & \ensuremath{-}0.01 [\ensuremath{-}0.18, 0.16] & no \\
Five-mode Fitts composite & N = 202 & normalised difference time & person & main & Algebraic re-expression & 0.09 & 1 & .764 & 1.000 & .793 & 0.03 [\ensuremath{-}0.15, 0.21] & no \\
Fitts task time: Head & N = 202 & normalised difference time & person & main & Algebraic re-expression & 0.09 & 1 & .761 & 1.000 & .993 & 0.02 [\ensuremath{-}0.13, 0.18] & no \\
Fitts task time: Eye & N = 202 & normalised difference time & person & main & Algebraic re-expression & 0.58 & 1 & .445 & 1.000 & .993 & 0.06 [\ensuremath{-}0.09, 0.20] & no \\
Fitts task time: Finger & N = 202 & normalised difference time & person & main & Algebraic re-expression & 0.02 & 1 & .879 & 1.000 & .993 & \ensuremath{-}0.01 [\ensuremath{-}0.16, 0.14] & no \\
Fitts task time: Controller raycast & N = 202 & normalised difference time & person & main & Algebraic re-expression & 0.28 & 1 & .595 & 1.000 & .993 & 0.04 [\ensuremath{-}0.12, 0.21] & no \\
Fitts task time: Controller direct & N = 202 & normalised difference time & person & main & Algebraic re-expression & 0.46 & 1 & .496 & 1.000 & .993 & \ensuremath{-}0.07 [\ensuremath{-}0.28, 0.14] & no \\
Difficulty-cell Fitts score & N = 202 & normalised difference time & person & main & Algebraic re-expression & 0.08 & 1 & .771 & 1.000 & .874 & 0.02 [\ensuremath{-}0.14, 0.19] & no \\
Digital literacy & N = 202 & A/B task time & part & main & Time & 6.15 & 1 & .013 & .052 & .026 & \ensuremath{-}0.12 [\ensuremath{-}0.21, \ensuremath{-}0.03] & no \\
Digital literacy & N = 202 & A/B task time & part & \ensuremath{\times} Part & Time & 2.22 & 1 & .136 & .136 & .136 & \ensuremath{-}0.05 [\ensuremath{-}0.10, 0.01] & no \\
Digital literacy & N = 202 & A/B errors & part & main & Errors & 4.92 & 1 & .026 & .079 & .052 & \ensuremath{-}0.19 [\ensuremath{-}0.35, \ensuremath{-}0.02] & no \\
Digital literacy & N = 202 & A/B errors & part & \ensuremath{\times} Part & Errors & 0.68 & 1 & .410 & .410 & .410 & \ensuremath{-}0.04 [\ensuremath{-}0.15, 0.06] & no \\
Digital literacy & N = 202 & A/B selection distance & part & main & Selection distance & 0.10 & 1 & .751 & 1.000 & .751 & \ensuremath{-}0.02 [\ensuremath{-}0.16, 0.12] & no \\
Digital literacy & N = 202 & A/B selection distance & part & \ensuremath{\times} Part & Selection distance & 0.03 & 1 & .852 & .852 & .852 & \ensuremath{-}0.01 [\ensuremath{-}0.08, 0.06] & no \\
Digital literacy & N = 202 & A/B adjusted time & part & main & Adjusted time & 7.23 & 1 & .007 & .022 & .011 & \ensuremath{-}0.13 [\ensuremath{-}0.22, \ensuremath{-}0.04] & yes \\
Digital literacy & N = 202 & A/B adjusted time & part & \ensuremath{\times} Part & Adjusted time & 2.12 & 1 & .145 & .145 & .145 & \ensuremath{-}0.04 [\ensuremath{-}0.10, 0.01] & no \\
Digital literacy & N = 202 & A/B error rate & part & main & Algebraic re-expression & 4.09 & 1 & .043 & .233 & .076 & \ensuremath{-}0.12 [\ensuremath{-}0.24, \ensuremath{-}0.00] & no \\
Digital literacy & N = 202 & A/B error rate & part & \ensuremath{\times} Part & Algebraic re-expression & 0.00 & 1 & .975 & 1.000 & .978 & 0.00 [\ensuremath{-}0.07, 0.08] & no \\
Digital literacy & N = 202 & A/B accuracy & part & main & Algebraic re-expression & 4.49 & 1 & .034 & .233 & .076 & 0.13 [0.01, 0.25] & no \\
Digital literacy & N = 202 & A/B accuracy & part & \ensuremath{\times} Part & Algebraic re-expression & 0.00 & 1 & .978 & 1.000 & .978 & 0.00 [\ensuremath{-}0.08, 0.08] & no \\
Digital literacy & N = 202 & total time & person & main & Time & 6.18 & 1 & .013 & .052 & .026 & \ensuremath{-}0.14 [\ensuremath{-}0.25, \ensuremath{-}0.03] & no \\
Digital literacy & N = 202 & total errors & person & main & Errors & 4.46 & 1 & .035 & .079 & .052 & \ensuremath{-}0.17 [\ensuremath{-}0.32, \ensuremath{-}0.01] & no \\
Digital literacy & N = 202 & total selection distance & person & main & Selection distance & 0.24 & 1 & .627 & 1.000 & .751 & \ensuremath{-}0.04 [\ensuremath{-}0.20, 0.12] & no \\
Digital literacy & N = 202 & total adjusted time & person & main & Adjusted time & 7.19 & 1 & .007 & .022 & .011 & \ensuremath{-}0.15 [\ensuremath{-}0.26, \ensuremath{-}0.04] & yes \\
Digital literacy & N = 202 & difference time & person & main & Time & 1.86 & 1 & .173 & .274 & .173 & 0.11 [\ensuremath{-}0.05, 0.26] & no \\
Digital literacy & N = 202 & difference errors & person & main & Errors & 0.43 & 1 & .514 & .514 & .514 & \ensuremath{-}0.05 [\ensuremath{-}0.22, 0.11] & no \\
Digital literacy & N = 202 & difference selection distance & person & main & Selection distance & 1.48 & 1 & .225 & .674 & .674 & 0.10 [\ensuremath{-}0.06, 0.26] & no \\
Digital literacy & N = 202 & difference adjusted time & person & main & Adjusted time & 1.49 & 1 & .223 & .223 & .223 & 0.10 [\ensuremath{-}0.06, 0.25] & no \\
Digital literacy & N = 202 & ratio time & person & main & Time & 2.21 & 1 & .137 & .274 & .173 & 0.12 [\ensuremath{-}0.04, 0.28] & no \\
Digital literacy & N = 202 & total error rate & person & main & Algebraic re-expression & 4.19 & 1 & .041 & .233 & .076 & \ensuremath{-}0.16 [\ensuremath{-}0.31, \ensuremath{-}0.01] & no \\
Digital literacy & N = 202 & total accuracy & person & main & Algebraic re-expression & 4.53 & 1 & .033 & .233 & .076 & 0.16 [0.01, 0.31] & no \\
Digital literacy & N = 202 & difference error rate & person & main & Algebraic re-expression & 0.42 & 1 & .515 & .947 & .515 & \ensuremath{-}0.05 [\ensuremath{-}0.22, 0.11] & no \\
Digital literacy & N = 202 & difference accuracy & person & main & Algebraic re-expression & 0.51 & 1 & .474 & .947 & .515 & 0.06 [\ensuremath{-}0.10, 0.22] & no \\
Digital literacy & N = 202 & normalised difference time & person & main & Algebraic re-expression & 2.21 & 1 & .137 & .411 & .192 & 0.12 [\ensuremath{-}0.04, 0.28] & no \\
\end{longtable}
\normalsize\noindent \textit{Note. Post hoc exploratory extension of the planned analyses (specified before the expanded models were run, after the planned analyses; not preregistered, not confirmatory). Outcomes = the 26 exported and derived TMT-VR variables. Outcomes other than the part-level and total wrong-target counts were normalised by ordered-quantile transform, standardised, and analysed with Gaussian linear mixed models (part level: Part fixed effect and participant random intercept) or Gaussian linear models (person level). Part-level and total wrong-target counts were analysed with negative-binomial (nbinom2) models, with a participant random intercept at the part level. The signed B \ensuremath{-} A error difference was normalised and analysed with a Gaussian linear model because it can take negative values. Fitts-derived predictors are adjusted for age; digital literacy is adjusted for age and education (all standardised). Test = main effect of the predictor or, at the part level, its interaction with Part. Holm and BH values are adjusted within family (predictor family \ensuremath{\times} outcome family \ensuremath{\times} test type; the five single-mode task times form one pooled family). Algebraic re-expression rows (error rate, correctness accuracy, normalised difference) are deterministic functions of the error and time variables and are never counted as independent evidence. b = standardised coefficient (SD per predictor SD) for Gaussian models and log rate ratio per predictor SD for negative-binomial count models. The fifteen Holm-significant rows are the completion-time and error-adjusted-time associations of the difficulty-cell Fitts score, the five-mode composite, the head-gaze and controller-direct task times, and the error-adjusted-time associations of digital literacy reported in manuscript Table 6.}